\documentclass[]{spie}  

\usepackage{amsmath,amsfonts,amssymb}
\usepackage{graphicx}
\usepackage[colorlinks=true, allcolors=blue]{hyperref}
\usepackage{booktabs}
\usepackage[absolute,overlay]{textpos}
\usepackage{xcolor}
\usepackage{fancyhdr}

\title{Efficient representation of 3D spatial data for defense-related applications}

\author[]{Benjamin Kahl}
\author[]{Marcus Hebel}
\author[]{Michael Arens}
\affil[]{Fraunhofer IOSB, Ettlingen, Fraunhofer Institute of Optronics, System Technologies and Image Exploitation. Fraunhofer Center for Machine Learning.\\Gutleuthausstr. 1, 76275 Ettlingen, Germany}

\authorinfo{Further author information:\\Benjamin Kahl: benjamin.kahl@iosb.fraunhofer.de, phone: +49 7243 992-393, ORCID: 0009-0009-3423-2823\\	Marcus Hebel: marcus.hebel@iosb.fraunhofer.de, phone: +49 7243 992-323, ORCID: 0000-0003-4301-7286\\    Michael Arens: michael.arens@iosb.fraunhofer.de, phone: +49 7243 992-147, ORCID: 0000-0002-7857-0332}

\begin{document}

\begin{textblock*}{17.5cm}(2cm, 0.3cm) 
    \noindent
    \small \textcolor{gray}{\fontfamily{qcr}\selectfont This is an author-prepared version. The original publication can be found in the SPIE Digital Library: Benjamin Kahl, Marcus Hebel, and Michael Arens "Efficient representation of 3D spatial data for defense-related applications", Proc. SPIE 13679, Artificial Intelligence for Security and Defence Applications III, 1367911 (27 October 2025); https://doi.org/10.1117/12.3069693
    Systematic reproduction and distribution, duplication of any material in this paper for a fee or for commercial purposes, or modification of the content of the paper are prohibited.}
\end{textblock*}

\begin{textblock*}{13.5cm}(4cm, 24.0cm) 
    \centering
    \fontfamily{phv}\selectfont Artificial Intelligence for Security and Defence Applications III, edited by Hugo J. Kuijf,
Radhakrishna Prabhu, Yitzhak Yitzhaky, Proc. of SPIE Vol. 13679, 1367911
2025 Published by SPIE · 0277-786X · doi: 10.1117/12.3069693
\end{textblock*}

\maketitle

\begin{abstract}
Geospatial sensor data is essential for modern defense and security, offering indispensable 3D information for situational awareness. This data, gathered from sources like lidar sensors and optical cameras, allows for the creation of detailed models of operational environments.

In this paper, we provide a comparative analysis of traditional representation methods, such as point clouds, voxel grids, and triangle meshes, alongside modern neural and implicit techniques like Neural Radiance Fields (NeRFs) and 3D Gaussian Splatting (3DGS). Our evaluation reveals a fundamental trade-off: traditional models offer robust geometric accuracy ideal for functional tasks like line-of-sight analysis and physics simulations, while modern methods excel at producing high-fidelity, photorealistic visuals but often lack geometric reliability.

Based on these findings, we conclude that a hybrid approach is the most promising path forward. We propose a system architecture that combines a traditional mesh scaffold for geometric integrity with a neural representation like 3DGS for visual detail, managed within a hierarchical scene structure to ensure scalability and performance.
\end{abstract}


\section{Introduction}
\label{sec:intro}  

A critical differentiator in modern defense is the ability to perceive and act upon information faster and more accurately than an adversary, a trend underscored by the rise of open source and social media intelligence \cite{osint_in_military, social_media_analytics_mdpi}. While many battle management systems still rely on traditional 2D interfaces that can limit spatial awareness, emerging systems are increasingly incorporating algorithms from computer graphics and AI to build richer models of the battlespace \cite{ai_for_c2}.

\begin{figure} [ht]
   \begin{center}
   \begin{tabular}{c}
   \includegraphics[height=6cm]{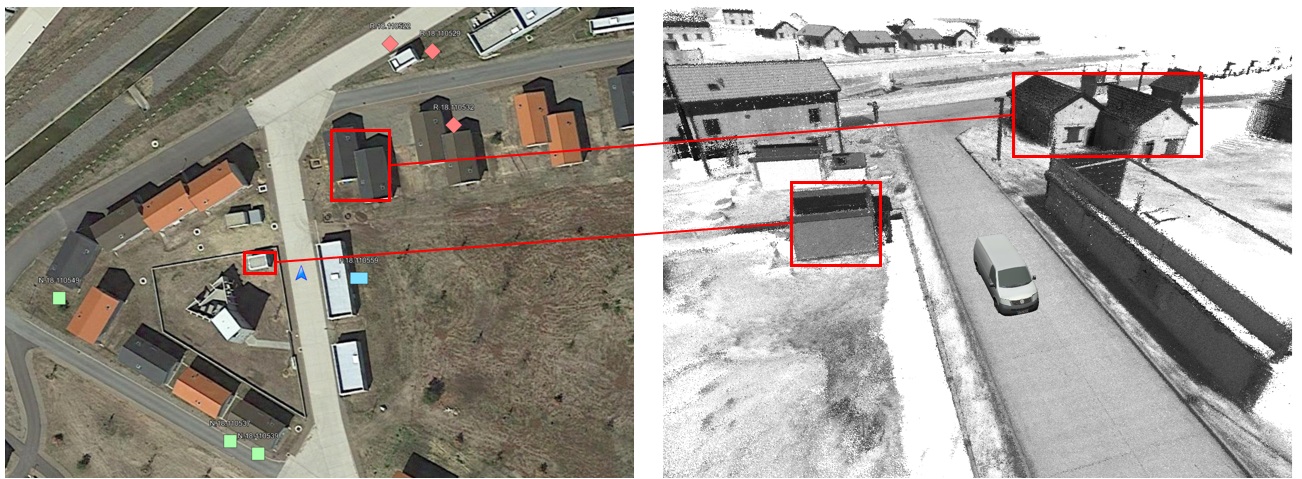}
   \end{tabular}
   \end{center}
   \caption[example] 
   { \label{fig:tak_v_lid} 
   Screenshot from the popular geospatial information software WinTAK\cite{Wintak} (left), and ambient-light lidar-scan of the same area (right). A 3D representation can provide richer information and enhance decision-making. 
   }
\end{figure}

A well-integrated 3D operational picture offers considerable benefits over 2D counterparts, enabling more intuitive analysis for tasks like route optimization, line-of-sight assessments, and terrain masking, as can be seen in Fig. \ref{fig:tak_v_lid}. The primary technical challenge, however, lies in integrating, processing, and updating massive, heterogeneous data sets from disparate sensors in real-time. Raw sensor data is often too voluminous or unstructured for immediate tactical use, making its transformation into a unified model a significant hurdle.

This paper examines the key stages of the data pipeline for creating such operational models. We analyze common data sources, formats, and models and compare approaches for transforming raw data into coherent, actionable 3D tactical representations. The goal is to identify best practices and future directions for enabling effective decision-making through advanced digital battlespace visualization.

\subsection{Previous Work}

The digital representation of 3D scenes has evolved from traditional explicit methods like \textit{point clouds}, \textit{voxel grids}, and \textit{meshes} to powerful implicit and neural representations.

Machine learning innovations have introduced powerful implicit and neural representations that can capture complex scenes with remarkable fidelity from sensor data. The concept of representing geometry implicitly through a learned continuous function was notably advanced by Park et al. (2019) with \textit{DeepSDF} \cite{DeepSDF}, which leverages deep-learning techniques to encode a shape's signed distance function into a \textit{neural field}.

This idea was later expanded upon by Mildenhall et al. through the introduction of \textit{Neural Radiance Fields} (NeRF) \cite{Nerf}, which encode a scene's direction-based radiance. A concept which has been adapted and implemented into standardized tools, such as the notable \textit{Nerfstudio} by Tancik et al. in 2022 \cite{nerfstudio}.

In 2023, Kerbl et al. \cite{gaussian_splatting} introduced \textit{3D Gaussian Splatting} (3DGS), a method that achieves real-time rendering of high-fidelity scenes by representing them as a set of optimized 3D Gaussians.

Many of these techniques have already been extensively compared and analyzed. A 2024 survey by Wang \cite{3d_repr_survey} provides a qualitative assessment of these various representation forms, from classic meshes to modern innovations like NeRFs and Gaussian Splatting, and lists notable datasets for each one. Similarly, other surveys have focused on more specific applications of machine learning to 3D data. For instance, surveys by Shi et al. (2023) \cite{3d_gen_models_survey} and Li et al. (2024) \cite{advances_3d_gen_survey} explore the burgeoning field of generative models and their transfer from 2D to 3D representations. Earlier, a survey by Ahmed et al. (2019) \cite{survey_deeplearning_3d_repr_survey} reviewed the use of other deep learning techniques, such as segmentation and recognition, on various 3D data representations.

In this paper, we assess many of the same representation forms, but remain focused on their applicability for use-cases beyond rendering. Subsequently, we will propose a hybrid solution that addresses their shortcomings.

\section{Overview}

The flow of information within Command-and-Control (C2) systems is commonly separated into an \textit{OODA-loop} (observe, orient, direct and act) for decision making, and an \textit{Intelligence Cycle} (IC) for assessments \cite{springer_intelligence_cycle}. ICs consist of various continuous stages that typically involve a stage of \textit{directing} and planning a data collection effort, the \textit{collection} of data itself, the subsequent \textit{processing} of said data into a congruent package and the \textit{dissemination} of that package to the relevant end users \cite{springer_intelligence_cycle}.

For the purposes of this paper, we restrict our scope to the stage of processing all available data into a digestible representation form and the underlying methodology thereof.

\begin{figure} [ht]
   \begin{center}
   \begin{tabular}{c}
   \includegraphics[height=5cm]{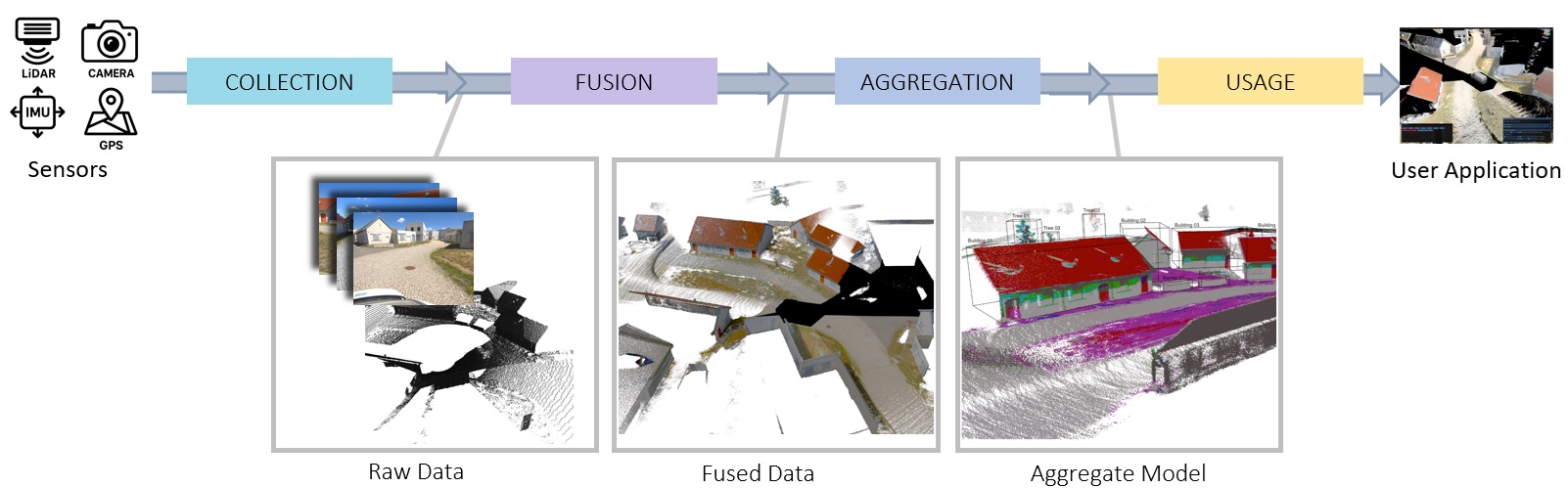}
   \end{tabular}
   \end{center}
   \caption[example] 
   { \label{fig:g3} 
   Generalized steps of a sensor-to-representation pipeline. Raw sensor data is collected, fused, then an aggregate model is formed, which is then queried and visualized by the end-user application.
   }
\end{figure}

This process, which takes us from the raw data delivered by sensors to a usable, congruent application can broadly be subdivided into the four stages of \textit{Collection}, \textit{Fusion}, \textit{Aggregation} and \textit{Usage} (as depicted in Fig. \ref{fig:g3}). We will outline the processes of each stage below.

\subsection{Collection Stage}

Before a coherent 3D tactical map can be created, an acquisition of relevant data is necessary. Data can be acquired through different kinds of sensors mounted to various platforms such as UAVs, ground vehicles, fixed installations, satellites or otherwise. 

Despite the broad diversity of platforms, the sensors involved can generally be classified into a few functional categories \cite{mobile_mapping_sensors}:

\begin{itemize}

    \item \textbf{Camera Sensors} are passive sensors that capture 2D images by measuring electromagnetic radiation on a photosensor array. While most cameras exclusively produce imagery in the visible light spectrum, more specialized variants operate in other spectra, such as the infrared or ultraviolet. The resulting data is typically delivered as a 2D array or texture, often geotagged and timestamped.

    \item \textbf{Lidar Sensors} are active sensors that emit laser pulses and measure their return times to determine the distance to a reflecting surface. Common lidar architectures include a rotating array of emitters and sensors. Each rotation produces a data-set including a measured distance (as well as \textit{if} a distance was measured) for each pulse. Some lidar sensors also capture additional data, such as the reflectivity or ambient illumination of the surface (see Fig. \ref{fig:lida_data}).
    
    This data is often managed in the form of \textit{3D point clouds} containing all laser hits. However, it is important to note that, while point clouds are an effective way to represent solid surfaces, lidar systems also implicitly measure volumetric information of a \textit{negative space} of where lidar pulses passed through, confirming the space as empty.
    
    Depth-enabled cameras can serve a similar purpose to lidar sensors.

    \item \textbf{Geolocation and Positional Sensors} like \textit{intertial navigation systems} (INS) often combine measurements from multiple sub-sensors such as GNSS receivers, IMUs and odometry sensors to produce an estimate on the measurement platform's pose in some shared spatial reference frame. Raw data is often delivered in a position and rotation in WGS84 or ECEF coordinates alongside a timestamp. Without accurate positioning, even the most detailed image or point cloud has limited value.

    \item \textbf{Miscellaneous Environmental Sensors} include any auxiliary sensors that can provide contextual information about local conditions, such as temperature, humidity, atmospheric pressure, sound levels, or radiation. While these are not always essential for constructing 3D spatial models, they may be crucial for specific mission types or for enhancing situational awareness. These sensors usually produce scalar values tagged with a time.

\end{itemize}

\subsection{Fusion Stage}

\begin{figure} [ht]
   \begin{center}
   \begin{tabular}{c}
   \includegraphics[height=8.7cm]{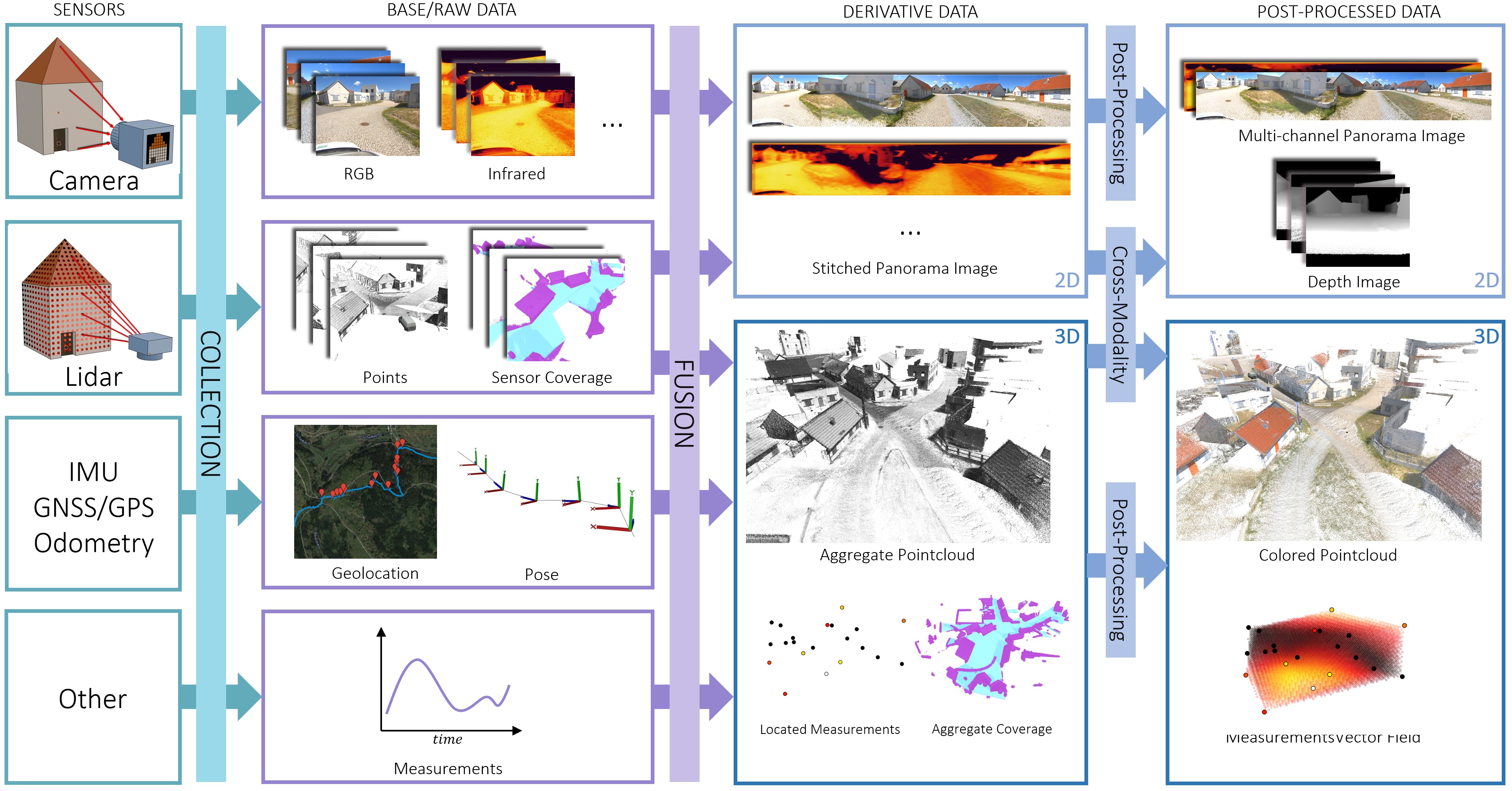}
   \end{tabular}
   \end{center}
   \caption[example] 
   { \label{fig:overview_stages} 
   Overview of the first two stages. Derivative data combines and filters raw data from multiple sensors.
   }
\end{figure}

The fusion stage is responsible for transforming heterogeneous sensor outputs into a unified spatiotemporal format. This involves aligning data across different coordinate systems, synchronizing timestamps, and applying necessary sensor-specific corrections. Effective fusion is a prerequisite for constructing a coherent model of the environment, as it allows data from disparate sources to be meaningfully combined (see Fig. \ref{fig:overview_stages}).

\subsubsection{Processing}


A fundamental aspect of sensor fusion is the normalization of positional and temporal references. Lidar measurements and camera images need to be lined up and converted into a consistent global or mission-specific coordinate space. Temporal synchronization is equally crucial; even small discrepancies in timestamp alignment can lead to significant registration errors, especially for fast-moving platforms.


With an adequate positional alignment, the data from miscellaneous environmental sensors, such as temperature, can be interpreted as a 3D vector field (or scalar field), where locations without measurements are interpolated to the nearest available values.

Camera images need to be undistorted, which effectively means removing any distortion coming from lenses and turning the image into something an equivalent pinhole camera would produce, where lines that are straight in reality also appear as straight lines in the image. This allows for an easier projection of points into the image via a simple frustum-projection, as well as other operations.
Furthermore, multiple camera images can be stitched to form a unified panoramic image.
If a camera's pose is unknown, \textit{structure-from-motion} (SfM) methods, such as \textit{COLMAP}\cite{colmap1, colmap2}, can be used to find out camera poses, as well as create additional point-clouds from feature-matching. 

\subsubsection{Post-Processing} 

Once sensor data is unified, it undergoes various post-processing operations. For camera images, this often includes anonymization or adjustments to brightness and contrast. For lidar data, common operations are outlier removal and motion filtering.

As shown in Fig. \ref{fig:overview_stages}, a key advantage is the ability to perform cross-modality enrichment, where data from one sensor enhances another. For example, camera images can be used to apply color to lidar point clouds, and point clouds can be used to generate depth images.

The final, fused output for each timestamp is a coherent data bundle that combines images (including stitched 360° views), lidar point clouds, and any additional measurements as vector fields. This comprehensive dataset serves as the foundation for creating coherent operational pictures.

\subsection{Aggregation Stage}

The datasets resulting from the fusion stage are detailed and cohesive, but too large to be queried, analyzed or visualized efficiently. Thus, an aggregate model needs to be constructed that captures essential spatial structures and semantics within a single structure,  while minimizing redundancy and resource demands.

We will outline and compare five different forms of representation in the next chapter and evaluate these for their individual strengths and weaknesses. Their adequacy is given by their performance based on the use-cases required by the last stage of our pipeline.

\subsection{Usage Stage}

The final stage of the tactical-map pipeline centers on the utilization of the aggregated scene representations. Once sensor data has been fused and compressed into a coherent model, its ultimate value is measured by how well it supports the needs of the end user.

\begin{figure} [ht]
   \begin{center}
   \begin{tabular}{c}
   \includegraphics[height=15cm]{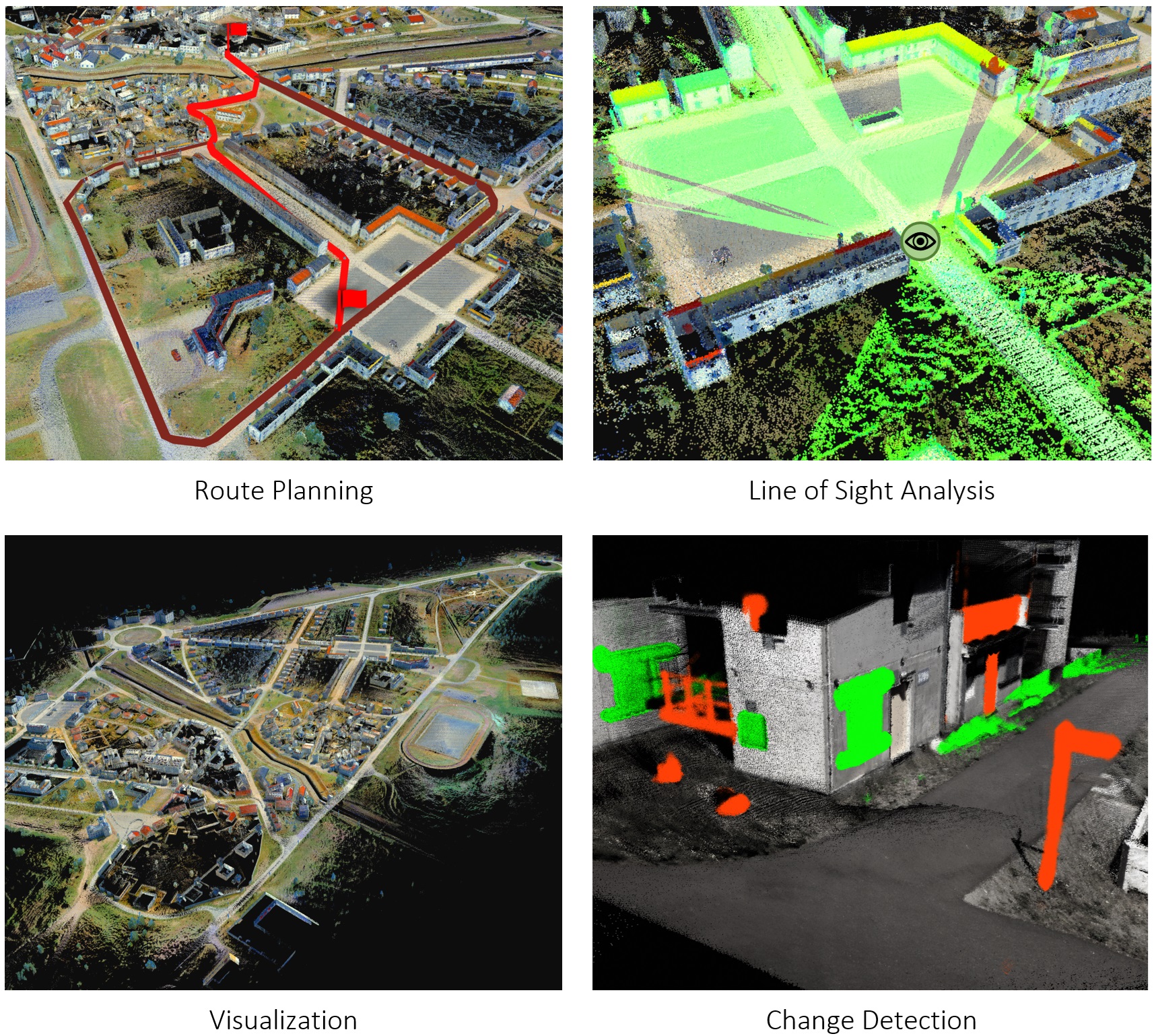}
   \end{tabular}
   \end{center}
   \caption[example] 
   { \label{fig:use-cases} 
   Examples of common use cases for a 3D representation model. 
   }
\end{figure}

The exact details and requirements depend heavily on the nature of the user application and it's purpose, but some common procedures include (see Fig. \ref{fig:use-cases}): 

\begin{itemize}
    \item \textbf{Line-of-sight analysis}: Assessing visibility from a given location, accounting for terrain, obstacles, and occlusions.
    
    \item \textbf{Visualization}: Rendering the environment in a way that is intuitive, photorealistic, or semantically informative for human operators or decision systems. The computational complexity of visualization and line-of-sight analysis are closely linked.
    
    \item \textbf{Route planning}: Identifying viable paths through the environment, often requiring integration with semantic information (e.g., traversable surfaces, hazards).

    \item \textbf{Physics/collision simulations}: Virtual environments can be leveraged to insert new objects into them and see how they interact with the surrounding environment.
    
    \item \textbf{Change detection and monitoring}: Comparing updated scans or mission data with existing maps to detect structural changes, new objects, or intrusions.
    
    \item \textbf{Simulation and synthetic data generation}: Using the map as a foundation for simulating agents, sensors, or physical phenomena in controlled environments.
\end{itemize}

\subsection{Evaluation Criteria}\label{eval_criteria}

Many of the computational processes required for the above listed use-cases rely on a similar set of operations. For the sake of our evaluation, we will regard the following:

\begin{itemize}
    \item \textbf{Write Performance}: The efficiency and ease of creating the representation from raw sensor data, as well as updating or modifying an existing model.

    \item \textbf{Memory Footprint}: The amount of memory required to store the representation, which dictates hardware constraints and scalability.

    \item \textbf{Fidelity}: The ability of the representation to accurately preserve different aspects of the source data. We distinguish between:
    \begin{itemize}
        \item \textbf{Surface Fidelity}: How accurately the model represents surface geometry, typically captured by lidar sensors.
        \item \textbf{Visual Fidelity}: How well the model preserves photorealistic appearance and texture information from camera sensors.
        \item \textbf{Volumetric Fidelity}: The capacity to represent non-surface data, such as atmospheric measurements or other phenomena within a volume.
    \end{itemize}

    \item \textbf{Computational Performance}: The efficiency of the representation when used for common downstream tasks, such as visibility estimation, rendering, route planning, or physics simulations.
\end{itemize}

In the following chapter, we assess how the various aggregation techniques perform across these dimensions. It's important to note that, in isolation, these criteria don't provide a complete picture on the advantages and disadvantages of each representation form.

\section{Evaluation of Aggregate Models}

\subsection{Point Clouds} \label{point_clouds}

\begin{figure} [ht]
   \begin{center}
   \begin{tabular}{c}
   \includegraphics[height=9cm]{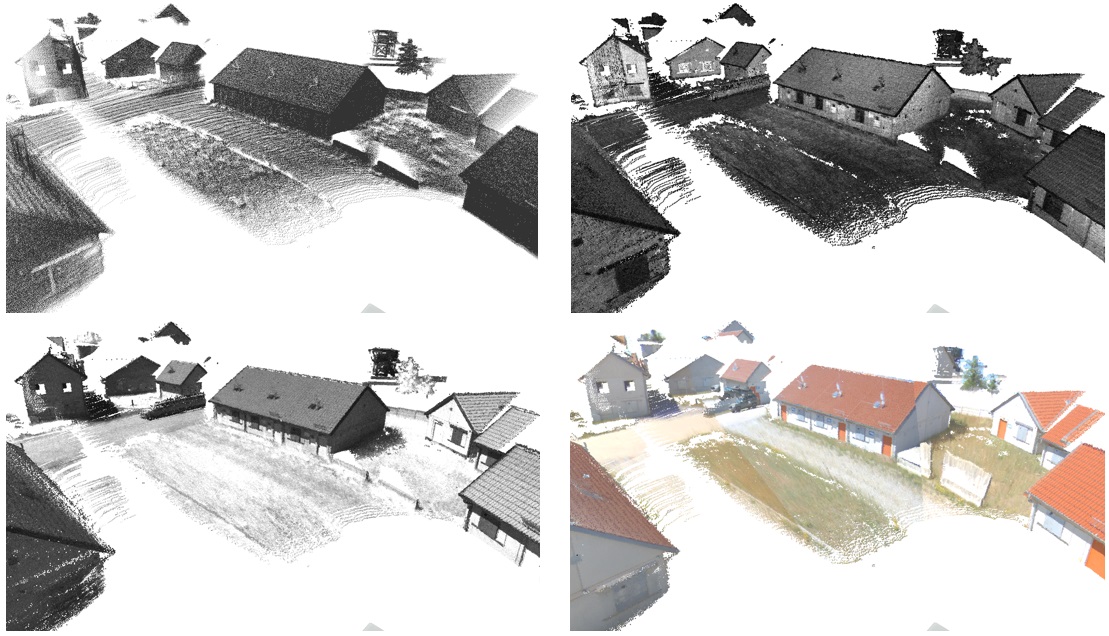}
   \end{tabular}
   \end{center}
   \caption[example] 
   { \label{fig:lida_data} 
   An example of a multi-layer point-cloud captured by a lidar sensor \cite{modissa}, consising of points (top-left), pulse-intensity (top-right), ambient light (bottom-left) and RGB colors captured by a complementary camera rig (bottom-right).
   }
\end{figure}

A \textit{point cloud} is a fundamental 3D representation consisting of a set of points $P = \{ p_i \mid p_i \in \mathbb{R}^3 \}_{i=1}^N$ in Euclidean space. Each point $p_i$ can be augmented with additional attributes, such as color, intensity, or semantic labels (see Fig. \ref{fig:lida_data}). As a minimally processed output from lidar sensors, point clouds offer a simple and detailed representation that remains faithful to the captured data. However, their data is often sparse, unstructured, and unordered, lacking any explicit connectivity information. This representation is also inherently limited; it captures only the surface points where lidar pulses terminate, discarding valuable volumetric data about scanned-through areas (free space) versus unscanned regions (unknown space), and often neglects rich contextual information available from co-located cameras.

Traditional algorithms for handling point clouds almost invariably rely on spatial acceleration structures, such as $k$-d trees or octrees, to enable efficient spatial queries \cite{potree, point_rendering_survey} like nearest-neighbor searches or ray intersections. Even with these structures, operations that depend on surface or volume information, such as line-of-sight analysis, remain inefficient without inferring additional structure. Similarly, rendering point clouds often relies on \textit{Level-of-Detail} (LOD) systems and custom compute shaders due to regular GPU pipelines not being optimized for point-rasterization \cite{point_rendering, point_rendering2}.

The unstructured nature of point clouds also poses a fundamental challenge for deep learning models, as the convolutional kernels used in standard CNNs are not applicable to unordered sets. A seminal work addressing this was PointNet \cite{pointnet}, which introduced a neural network architecture capable of processing raw point clouds directly. PointNet achieves permutation invariance by applying a shared MLP onto the input points independently and then pooling the resulting feature vectors into a single, global feature vector using a symmetrical max-pooling operation (see Fig. \ref{fig:pcl_super}). Its successor, PointNet++ \cite{pointnet_plusplus}, extended this concept by introducing a hierarchical architecture to capture local geometric features at multiple scales, effectively creating a multi-scale understanding of the point cloud's geometry. Despite these advances, extracting robust and meaningful features from sparse point data remains an active area of research \cite{3d_gen_models_survey}.

\subsubsection{Surfels}

By augmenting each point $p_i$ with a normal vector $n_i$, it can be promoted into a \textit{surfel}, $s_i = (p_i, n_i)$ \cite{surfels}. The key innovation is that the rendering process for these surfels can be made differentiable \cite{surfel_splatting, surfelnerf}. 

\begin{figure} [ht]
   \begin{center}
   \begin{tabular}{c}
   \includegraphics[height=5.8cm]{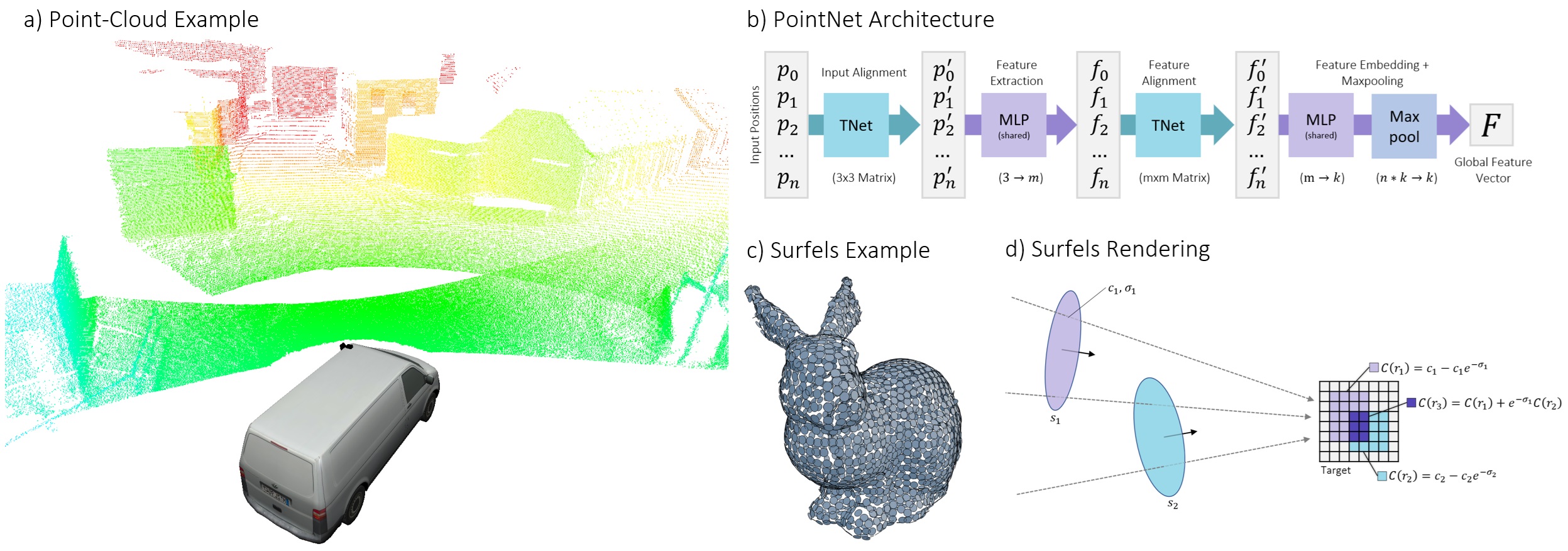}
   \end{tabular}
   \end{center}
   \caption[example] 
   { \label{fig:pcl_super} 
   a) Example of a point-cloud captured with a double-lidar setup using the MODISSA platform \cite{modissa}.
   b) The network architecture behind PointNet's pre-processing. Points undergo individual feature extraction. A set of learnable \textit{TNet} matrices ensure transformation invariance, while a symmetrical max-pooling layer ensures permutation invariance.
   c) A rendition of the \textit{Stanford Bunny} \cite{stanford} using surfels and
   d), An illustration of the differential rendering function of surfels \cite{surfelnerf}.
   }
\end{figure}

Given a depth-sorted set of surfels $\{s_i\}_{i=1}^N$ intersected by a pixel ray, the final color $C(x)$ of the pixel can be calculated through following rendering equation\cite{Nerf, surfelnerf}:

\begin{equation}
    C(r) = \sum_{i=1}^NT_i(1-\exp(-\sigma_i))c_i \quad
\end{equation}

where $c_i$ and $\sigma_i$ are the color (or radiance) and relative volume density of the respective surfels at their intersection points, and $T_i$ is the accumulated transmittance along the ray $T_i = exp(-\sum_{j=1}^{i-1}\sigma_j)$. An example of this process can be seen in Fig. \ref{fig:pcl_super}.

Because this blending process is differentiable, gradients can be backpropagated from a rendering loss (e.g., the difference between the rendered image and a ground-truth photo) to the surfel parameters. This enables gradient-based optimization to refine the 3D scene directly from 2D images \cite{surfelnerf, surfel_splatting}. This principle forms the theoretical basis for 3D Gaussian Splatting, which we evaluate in Section \ref{gaussian_splatting}. 

\subsection{Voxel Grids}

A voxel grid discretizes a continuous 3D space into a regular lattice of cubic elements, or voxels. Formally, a grid $G$ of resolution $N_x \times N_y \times N_z$ can be defined as a 3D tensor where each voxel $v_{i,j,k} \in G$ stores a particular value associated with its spatial location. This value can represent various properties, such as occupancy, color, or signed distance values. Like point clouds, voxel grids are trivial to construct or modify and are superb for storing volumetric information such as sensor coverage.

While primitive voxel grids have a cubic memory complexity ($O(N^3)$), in practice, most scenes are sparse, meaning the majority of voxels represent empty space. This observation motivates the use of sparse data structures such as octrees \cite{voxel_cone_tracing}. Although octrees can significantly alleviate memory requirements for sparse environments, their irregular, pointer-based structure disrupts the contiguous memory layout that makes dense grids computationally efficient \cite{VDBPaper}.

\begin{figure} [ht]
   \begin{center}
   \begin{tabular}{c}
   \includegraphics[height=11cm]{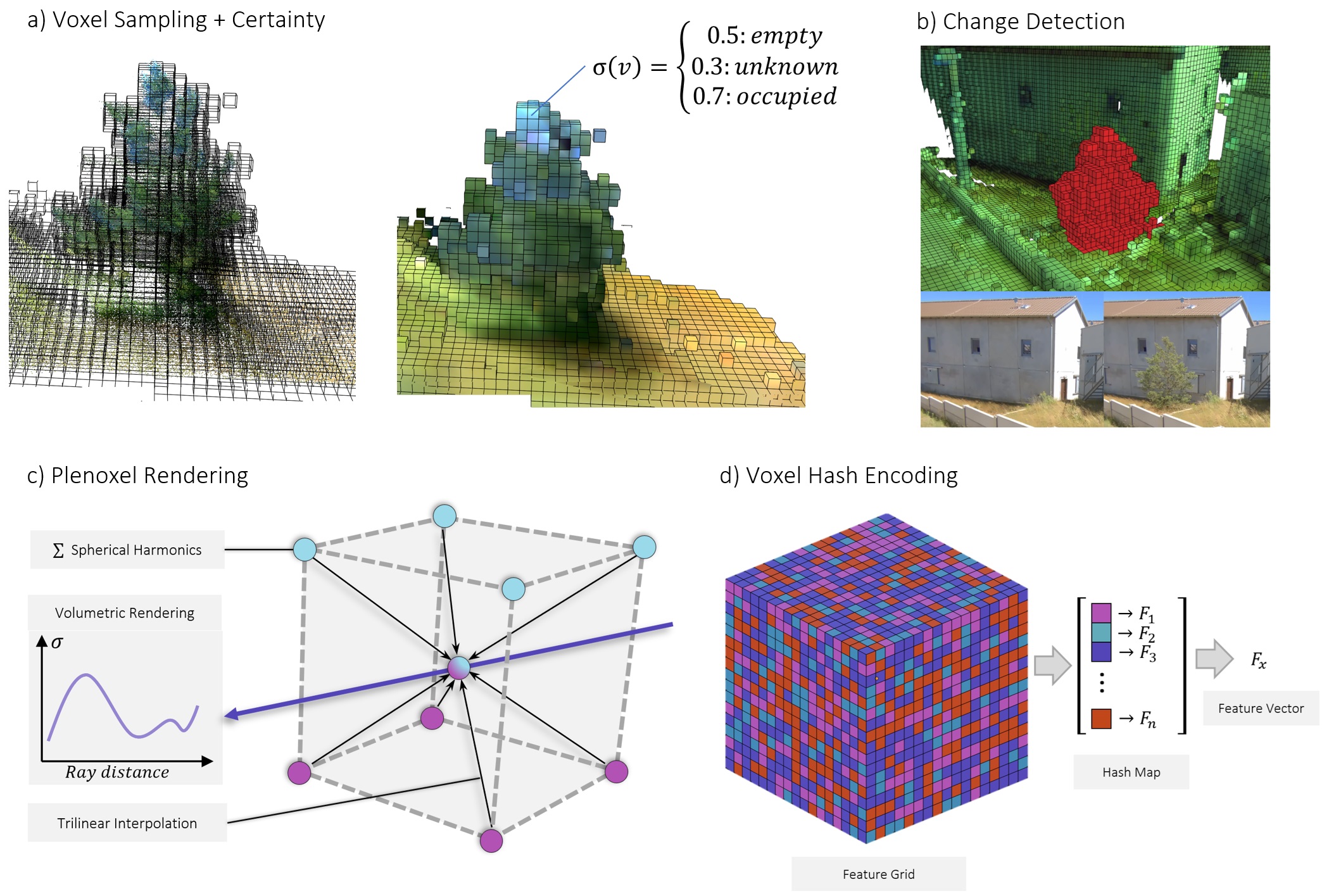}
   \end{tabular}
   \end{center}
   \caption[example] 
   { \label{fig:voxel_super} 
   a) An example of how voxel grids can be constructed from lidar point-clouds and sensor coverage encoded as a per-voxel certainty function.
   b) An example of how differences in this certainty function can easily be used to detect changes in the environment.
   c) Illustration of the differentiable, volumentric rendering function of plenoxels \cite{plenoxels}. A ray traversing a voxel yields a color and volume density value that is tri-linearly interpolated to the spherical harmonics on the edges of the voxel.
   d) The concept behind voxel hash encoding. Instead of containing the scene itself, the voxel map only points to learnable feature vectors in a hash-grid. Note that during read-out, the feature-vectors in the voxel-map are typically tri-linearly interpolated on the queried position.
   }
\end{figure} 

Depending on the acceleration structure used, the predictable, Euclidean structure of voxel grids allows for rapid neighborhood lookups, making them adequate for operations like 3D convolution. On the other hand, queries that require traversing the space, such as line-of-sight or ray casting, can be inefficient. A ray must be "marched" through the grid cell by cell, and the precision of any intersection test is fundamentally limited by the voxel size.

In the context of deep learning, their similarity to 2D images allowed for the straightforward application of 3D Convolutional Neural Networks (CNNs). Early works like DeepVoxels \cite{deepvoxels} for representation, 3D-R2N2 \cite{v3d_r2n2} for reconstruction and 3D-GAN \cite{3dgan} for shape generation demonstrated the viability of this approach. However, the aforementioned cubic memory complexity proved a significant hindrance on achievable resolution and training times \cite{3d_gen_models_survey}. As with other operations, acceleration structures like octrees can reduce this cost, but their non-regular structure does not lend itself well to processing by neural networks \cite{3d_gen_models_survey}.

Subsequent research to overcome these limitations diverged into two main paradigms: The first paradigm shifted focus from using grids as an input to a large network, to treating the grid itself as the set of optimizable parameters. A prominent example is Plenoxels \cite{plenoxels}, which utilizes a sparse voxel grid where each cell stores not only density but also coefficients for spherical harmonics (SH) to model view-dependent color. The entire set of parameters is then optimized directly against a collection of training images using a differentiable volumetric rendering process (see Fig. \ref{fig:voxel_super}c). This philosophy of directly optimizing explicit geometric primitives has been further advanced by methods like 3D Gaussian Splatting \cite{gaussian_splatting}, which replaces voxels with anisotropic 3D Gaussians (see Section \ref{gaussian_splatting}).

The second paradigm focused on using grids to accelerate implicit neural representations. Here, each grid cell holds a learnable, high-dimensional feature vector rather than a final value. To query a continuous point $\mathbf{p}$, the feature vectors from its neighboring voxels are interpolated, and this interpolated feature is processed by a very small multi-layer perceptron (MLP) to output the final density and color. This hybrid explicit-grid/implicit-network design was powerfully demonstrated by Instant-NGPs \cite{instant_ngps} (see Section \ref{instant-ngps}).

\subsection{Triangle Mesh}\label{meshes}

3D triangle meshes are among the most ubiquitous representations for 3D shapes. A mesh defines the surface of an object as a collection of vertices, edges, and polygonal faces. The vertices specify coordinates in 3D space, while the faces describe the surface connectivity between them. The widespread adoption of meshes in computer graphics has resulted in a mature ecosystem of tools, algorithms, and hardware acceleration, making them a compelling representation for virtual environments.

Modern Graphics Processing Units (GPUs) are highly optimized for rasterizing extremely large meshes and can perform on-the-fly operations like morphing and tessellation in real-time. This hardware support facilitates efficient computational operations. For instance, visibility analysis is handled through established techniques like $z$-buffering and shadow mapping, which can be further accelerated by dedicated ray-tracing hardware on modern GPUs \cite{rtx_radiosity}. Furthermore, the application of 2D textures allows meshes to achieve a high degree of visual fidelity at a very low memory cost.

\begin{figure} [ht]
   \begin{center}
   \begin{tabular}{c}
   \includegraphics[height=4.7cm]{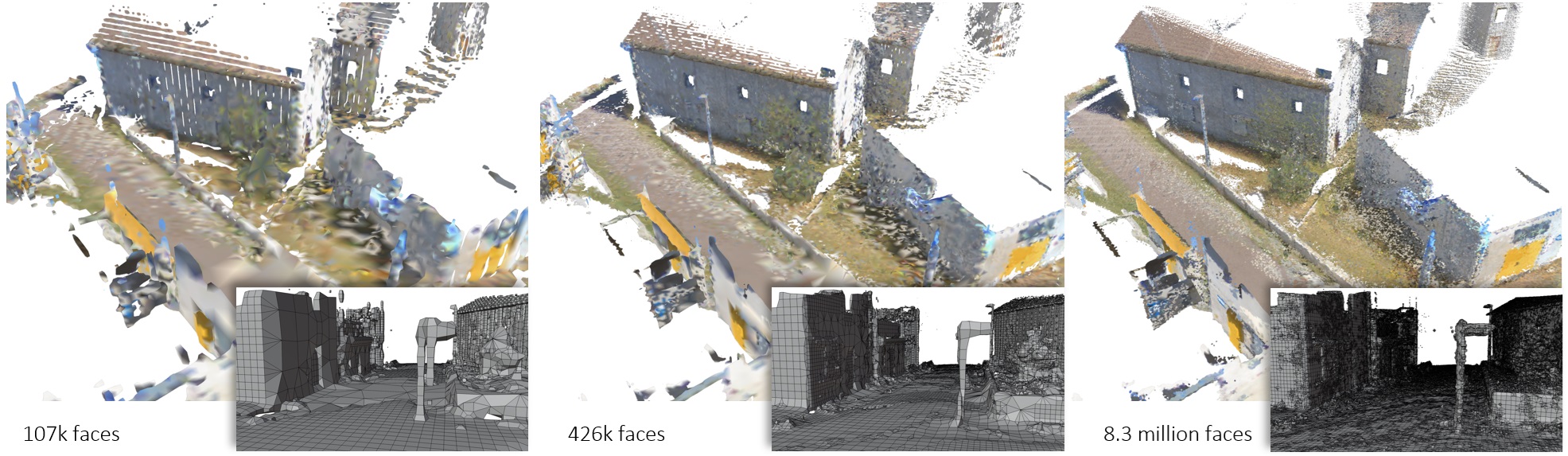}
   \end{tabular}
   \end{center}
   \caption[example] 
   { \label{fig:lod_overview} 
   Mesh of the same point-cloud with three different LODs (left one being lowest, right one being highest) and their respective polygon-counts. In this example, mesh coloring is performed on a per-vertex level, and could be significantly improved through the use of textures.
   }
\end{figure} 

Meshes do, however, lack volumetric expressiveness and can only describe the boundary of a shape. This limits them to binary volumetric distinctions (inside vs. outside) and prevents the representation of smooth interior properties that would be achievable with voxel grids or neural fields. Additionally, reliably constructing a topologically correct and accurate mesh from raw sensor data is a non-trivial challenge, often complicated by noise and incomplete data \cite{surface_reconstruction_survey}.

Applying deep learning methods to 3D meshes has historically been difficult due to their irregular, non-Euclidean structure. Unlike images, which are organized in a regular pixel grid, meshes have complex and variable vertex connectivity. Consequently, standard models like \textit{Convolutional Neural Networks} (CNNs), designed for grid-like data, cannot be directly applied \cite{3d_gen_models_survey}. Generating a new mesh is also complex, as it requires synthesizing not just vertex positions but also a plausible topological structure for the faces \cite{3d_gen_models_survey}.

To bridge this gap, differentiable rendering techniques have been developed. These methods enable a mesh to "learn" its own shape by comparing its rendered output to a set of target images, in much a similar fashion as surfels, plenoxels or Gaussian Splatting. Among the most notable implementations is the \textit{Neural 3D Mesh Renderer} by Kato et al. \cite{neural_mesh_renderer}. It uses standard, non-differentiable rasterization for the forward pass, but employs a proxy function in the backward pass to approximate the gradient of the rasterization step. This approximation is often sufficient to guide the optimization of the mesh's geometry. In contrast, the \textit{Soft Rasterizer} by Liu et al. \cite{softrasterizer1, softrasterizer2} introduces a fully differentiable rendering pipeline. It reformulates rasterization as a probabilistic process where every triangle contributes to the final color of each pixel, enabling a more direct and accurate flow of gradients.

Beyond mesh generation and optimization, other methods focus on applying deep learning directly to the analysis of existing mesh structures. A seminal work in this area is \textit{MeshCNN} by Hanocka et al. \cite{meshcnn}. MeshCNN allows certain meshes to be used as inputs for traditional CNNs by treating the geometric properties of the meshes edges analogously to pixels on an image and performing convolutions on rings of four neighboring edges in an order consistent with the respective face normal \cite{meshcnn}. This allows fully convolution tasks, such as segmentation, to be performed with traditional CNNs directly on meshes.

\subsection{Neural Fields}

At their simplest level, neural fields are multilayer perceptrons (MLPs, often with fully connected layers and no external memory) that take spatial coordinates as an input and output a value associated with that location. The concept was popularized by Mescheder et al. in 2019 with \textit{Occupancy Networks} \cite{occupancy_networks}, where a 3D shape is represented by an MLP that learns its occupancy function (i.e., whether a point is inside or outside the shape) based on input coordinates. Park et al. later expanded this concept with \textit{DeepSDF} \cite{deep_sdf}, which improves rendering performance and enables the inference of surface normals by encoding the space as a \textit{signed distance function} (SDF), which represents the distance to the nearest surface for every point in the represented space.

However, these early methods suffered from a significant \textit{spectral bias}, meaning they were inherently biased towards learning low-frequency functions and struggled to represent fine, high-frequency details like sharp edges, textures, or intricate geometry. The breakthrough addressing this issue came when Mildenhall et al. introduced \textit{Neural Radiance Fields} (NeRFs) \cite{Nerf}. With NeRF, they borrowed a concept first introduced by Vaswani et al. \cite{attention_nlp} in the realm of natural language processing and applied it to the completely different domain of 3D representations. This concept, known as \textit{positional encoding}, maps the low-dimensional input coordinates into a higher-dimensional feature space, making it easier for the network to learn high-frequency variations.

The specific implementation in NeRF is a deterministic mapping that creates a set of axis-aligned, exponentially spaced frequency features. A point $p=(x,y,z)$ in 3D space is transformed into a higher-dimensional point by applying the following function to each component of the coordinate vector:

\begin{equation}
\gamma(p)=(sin(2^0\pi p),cos(2^0\pi p),...,sin(2^{L-1}\pi p),cos(2^{L-1}\pi p))
\end{equation}

Where L is the number of chosen frequencies. The final vector is the concatenation of these features for $x$, $y$, and $z$, resulting in a $(3*2L)$-dimensional vector. This mapping allows the MLP to more easily represent high-frequency functions. Despite being popularized by NeRF, this technique is broadly applicable to improve detail in any neural field.

\begin{figure} [ht]
   \begin{center}
   \begin{tabular}{c}
   \includegraphics[height=15cm]{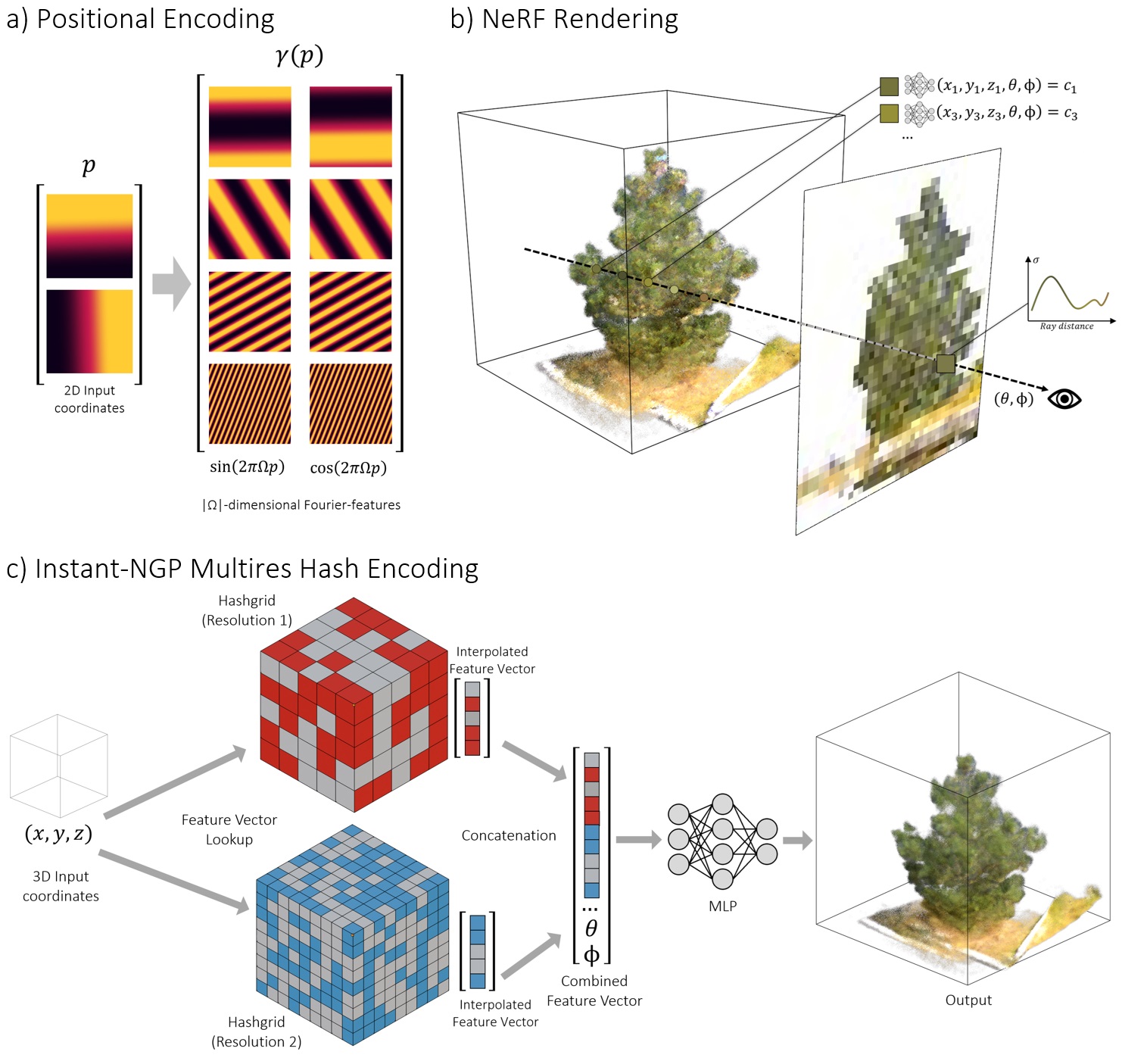}
   \end{tabular}
   \end{center}
   \caption[example] 
   { \label{fig:neural_fields_super} 
   a) The concept behind \textit{Random Fourier Features} (RFF) in 2D: The input coordinates are decomposed into a set of fourier features with randomized frequencies and directions.
   b) The differentiable rendering process of NeRFs: A ray is marched through the field, accumulating transmittance. For each sampling point, an inference of the network is performed.
   c) The concept behind multi-resolution hash encoding used in Instant-NGPs: The input coordinates sample a number of hashgrids with different resolutions for feature vectors that are tri-linearly interpolated for each grid, then concatenated alongside the view direction. A small MLP interprets the combined feature vector to output the radiance.
   }
\end{figure}

A more general approach, analyzed in detail by Tancik et al. \cite{fourier_rff}, uses \textit{Random Fourier Features} (RFF, see Fig. \ref{fig:neural_fields_super}a). This method relies on a matrix $\Omega$ of randomly sampled frequency vectors, where each vector defines both a direction and frequency:

\begin{equation}
\gamma(p)=(cos(2\pi\Omega p),sin(2\pi\Omega p))
\end{equation}

While the RFF approach is theoretically powerful, the original NeRF paper found great success with the first, simpler method using a fixed set of axis-aligned frequencies.

With these tools, neural fields have proven to be a powerful, continuous representation that can capture complex volumetric data within a memory-efficient network. The optimization process naturally encourages the network to use its limited capacity to represent the most important aspects of the environment.

However, neural fields are not trivial to train on raw sensor data, as the loss function often requires sampling points in known empty space, not just on measured surfaces. Furthermore, they are typically cumbersome to modify. A "targeted" update to a small area is difficult, as a local change can affect the global representation, usually requiring a full re-training of the network.

Similarly, performing traditional operations such as visibility determination or physics simulations on a neural field is computationally expensive, as these operations require repeatedly querying the network by ray-marching through the represented volume.

\subsubsection{Neural Radiance Fields (NeRFs)}

In addition to the aforementioned positional encoding, Mildenhall et al.'s NeRFs distinguished themselves from regular Neural Fields by introducing the viewing direction to the input of the network \cite{Nerf}. Instead of just a position $(x, y, z)$, a NeRF takes two angles encoding the viewing direction $(x, y, z, \theta, \phi)$ for inputs. Thus, the network produces not color, but a view-dependent \textit{radiance} $c_r$ and volume density $\omega_r$, and that allows the representation to mimic phenomena such as specular reflections or refractions. The network thus acts as a function of five parameters and four outputs:

\begin{equation}
M(x, y, z, \theta, \phi) \longrightarrow (r, g, b, \sigma)
\end{equation}

The rendering of a NeRF image is performed using classical volume rendering (see Fig. \ref{fig:neural_fields_super}b). For each pixel, a camera ray $r$ (given by $\theta$ and $\phi$) is marched through the scene, and the MLP is queried at $N$ sample points along it. The final color $C(r)$ is then computed by integrating the color and density values:

\begin{equation}
C(r) = \sum_{i=0}^{N} T_{i,r} c_{i,r} (1-exp(-\sigma_{i,r} \delta_{i}))
\end{equation}

where $c_{i,r}$ and $\sigma_{i,r}$ are the radiance and volume density at point $i$ in direction $r$, given by the output of the network, $\delta_{i}$ is the distance to the next sample point and $T_{i,r}$ is the transmittance, which tracks the accumulated occlusion of the ray through its sample points, given by $T_{i,r} = exp(-\sum_{j=1}^{i-1}\sigma_{j,r} \delta_{j})$.

Like with plenoxels or surfels, this process is differentiable, meaning the difference between a rendered image and its original counterpart can be backpropagated all the way back through the volume rendering formula. This allows NeRFs to be trained directly on a collection of camera images, enabling them to capture photorealistic detail far beyond what could be reconstructed from geometric data like a lidar point cloud alone.

With these adaptations, NeRFs produce highly photorealistic results and have become a foundational technique in 3D computer graphics \cite{3d_repr_survey}.

Despite the impressive visual quality, the original NeRF architecture does not solve many of the drawbacks inherent to neural fields. Namely, training times being extremely slow, targeted modifications being difficult, and because the geometry is stored implicitly, querying the scene for tasks like collision detection or physics simulations remains prohibitively expensive.

\subsubsection{Instant-NGPs}
\label{instant-ngps}

The slow training times of NeRFs were ameliorated when Müller et al. introduced \textit{Instant-NGPs} (Neural Graphics Primitives with a Multiresolution Hash Encoding) \cite{instant_ngps}. Instant-NGPs employ a hybrid explicit-implicit representation with a learnable multi-resolution hash grid that maps input coordinates to sets of rich, multi-dimensional feature vectors that encode features on various resolutions (see Fig. \ref{fig:neural_fields_super}c). 

Given a position $p=(x,y,z)$, the explicit multi-resolution hash-grid (effectively a series of voxel maps pointing to a limited set of learnable vectors), produces a combined set of vectors for each of the $N$ resolutions:

\begin{equation}
H(x, y, z) \longrightarrow (F_0, F_1, ... F_N)
\end{equation}

This combined feature vector, which now encodes multi-scale information about that point in space, is passed to a small MLP. The viewing direction $(\theta, \phi)$ is encoded separately and concatenated to an intermediate feature layer within the MLP, just before the color is predicted. The network then produces the final radiance and volume density:

\begin{equation}
M(F_0, F_1, ... F_N, \theta, \phi) \longrightarrow (r, g, b, \sigma)
\end{equation}

By offloading the bulk of the scene representation from the slow, large MLP to the fast, explicit hash grid, the network's task is simplified from representing the entire scene to just interpreting the rich features from the grid. This allows for a much smaller MLP, which dramatically reduces training and inference times from hours to mere seconds or minutes, all while producing results comparable to regular NeRFs.

While the improved training times make creation and modification operations more feasible, Instant-NGPs are still ill-suited as representations for operations requiring many visibility determinations or collision/physics simulations.

\subsection{Gaussian Splatting}\label{gaussian_splatting}

3D Gaussian Splatting (3DGS), introduced by Kerbl et al. in 2023 \cite{gaussian_splatting}, is the newest major evolution in the line of self-learning 3D representations. While NeRFs or Instant-NGPs store the scene representation implicitly within the weights of their network, 3DGS is an explicit representation, where the trained parameters directly define a set of discrete objects that make up the rendered geometry.

Similarly to \textit{surfels} (see Section \ref{point_clouds}), a 3DGS representation consists of a set of primitives called Gaussians which take the shape of anisotropic, 3-dimensional ellipsoids that have a position, covariance (shape and scale), color and opacity. In more advanced implementations they are also outfitted with spherical harmonics parameters, allowing them to adopt different colors when viewed from different directions.

A Gaussian $G_i$ at position $\mu_i$ and covariance matrix $\Sigma_i$ is given by the following equation:

\begin{equation}
G_i(p) = \exp\left(-\frac{1}{2}(p - \mu_i)^T \Sigma_i^{-1} (p - \mu_i)\right)
\end{equation}

\begin{figure} [ht]
   \begin{center}
   \begin{tabular}{c}
   \includegraphics[height=5.5cm]{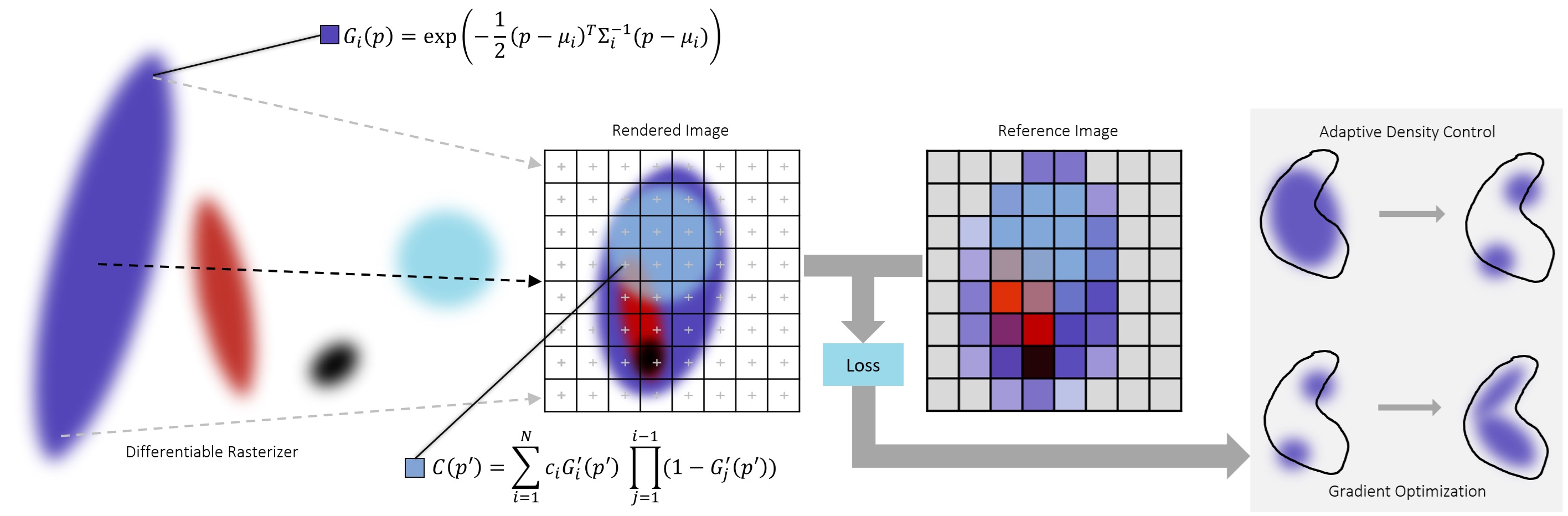}
   \end{tabular}
   \end{center}
   \caption[example] 
   { \label{fig:gaussian_splat_super} 
   The process of training a 3DGS representation: 3D Gaussians are rasterized ("splatted") onto a camera with the same perspective as a reference image. The loss for each tile/pixel is then back-propagated through the projection to the Gaussians' parameters, as well as towards adaptive density control.
   }
\end{figure}

In contrast to NeRFs, the rendering (ie. "splatting") of Gaussians does not require ray-marching and repeated sampling. While exact implementation details can differ, the most common approach involves sorting the Gaussians back-to-front and then performing a simple alpha-blend \cite{gaussian_splatting}. The final color of a pixel affected by $N$ Gaussians (sorted by depth) is given by

\begin{equation}
C(p') = \sum_{i=1}^{N} c_i G_i'(p') \prod_{j=1}^{i-1} (1 - G_j'(p'))
\end{equation}

where $c_i$ is the color of the $i$'th Gaussian and $G_i'$ is the 2D projection of the Gaussian onto the viewing plane (see Fig. \ref{fig:gaussian_splat_super}). This entire process is differentiable, allowing the properties of the Gaussians to be optimized directly to match a set of training images.

\begin{figure} [ht]
   \begin{center}
   \begin{tabular}{c}
   \includegraphics[height=4cm]{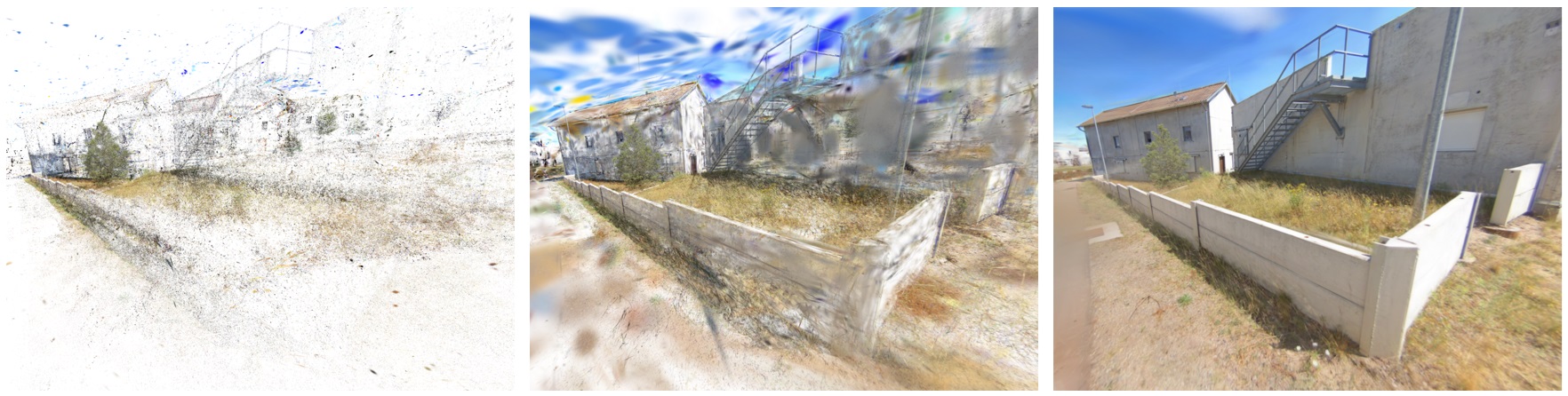}
   \end{tabular}
   \end{center}
   \caption[example] 
   { \label{fig:gs_splat_scaling} 
   A 3DGS render with the Gaussians scaled to 1\% (left), 40\% (center), and 100\% (right) of their size respectively. Unlike with point clouds, 3DGS only allocates a higher amount of Gaussians in locations with a coarse texture, acting as a kind of "compressed point cloud".
   }
\end{figure} 

In their implementation, Kerbl et al. use a Structure-from-Motion process to initialize the Gaussians from a sparse point cloud. Throughout training, these Gaussians are optimized using a process called \textit{adaptive density control}, which can split, clone, or prune Gaussians to better represent the scene's geometry.

The results are very impressive (see Fig. \ref{fig:gs_splat_scaling}). Renders of trained 3DGS representations are as photorealistic as NeRFs, but can be generated in real-time. The explicit model can also be expanded, modified, or selectively reduced with greater ease. Further improvements to this method have introduced distance-based LOD functionality or anti-aliasing, allowing for Gaussian Splatting to be used even on very large, complex environments \cite{gaussian_lod, gaussian_lod2}.

\begin{figure} [ht]
   \begin{center}
   \begin{tabular}{c}
   \includegraphics[height=10.5cm]{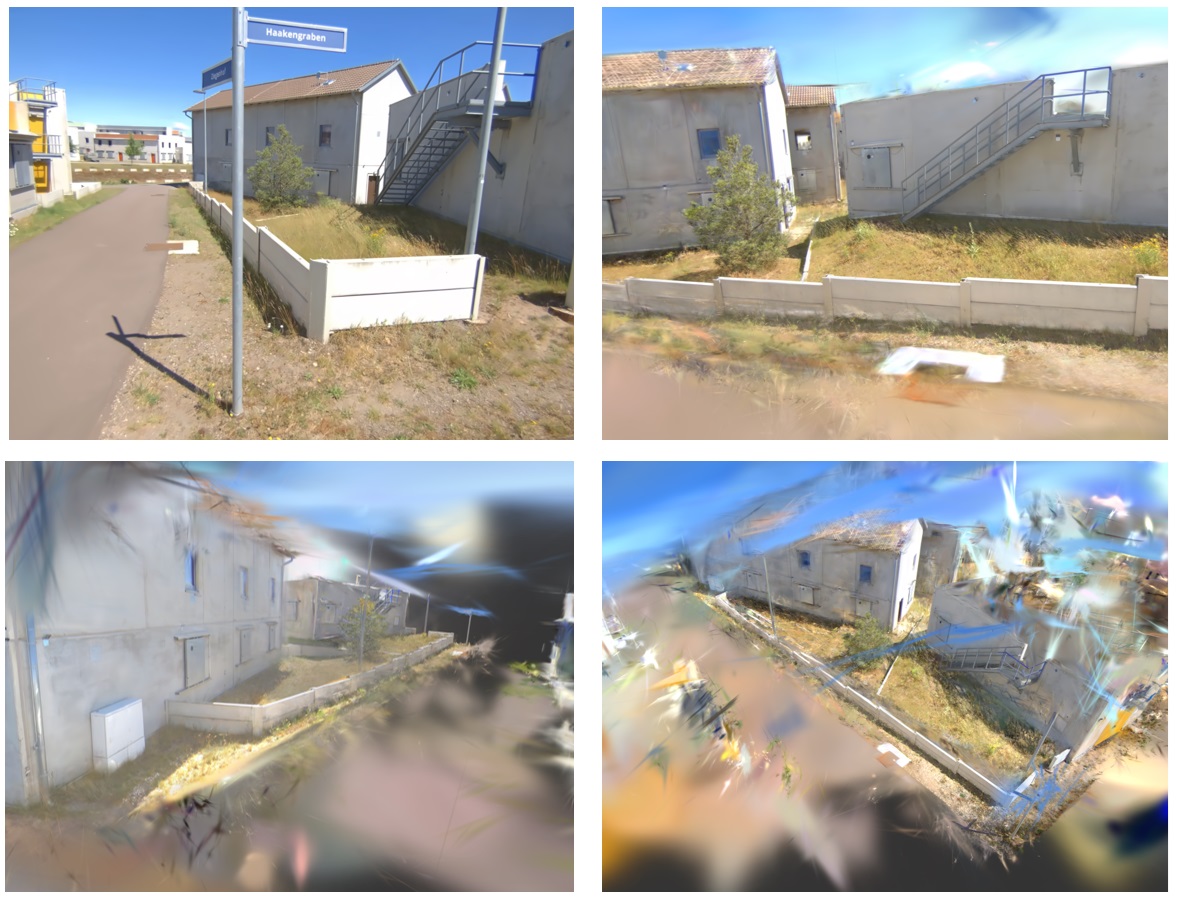}
   \end{tabular}
   \end{center}
   \caption[example] 
   { \label{fig:3dgs_examples} 
   The same 3DGS representation rendered from different camera angles. Angles that are similar (top right) or the same (top left) to the camera poses of the reference/training images are highly convicing. Diverging from the angles of the training data (bottom) showcases the unreliable geometric properties of 3DGS. Bottom left shows the road being partially transparent due to no reference image existing along that direction. Bottom right shows how coarse patterns in the sky and ground are being mimicked through Gaussians that don't accurately reflect the real surfaces.
   }
\end{figure}

Despite 3DGS' superb visual quality, its primary weakness stems from the fact that the optimization is driven purely by 2D image reconstruction. This can lead to geometrically inaccurate or "hollow" representations that look correct from the training viewpoints but may not reflect the true underlying scene structure \cite{sugar} (see Fig. \ref{fig:3dgs_examples}). Consequently, operations that rely on accurate geometry, such as route planning, physics simulations, or visibility calculations, can yield unreliable results. Furthermore, volumetric information, such as measurement coverage, cannot be encoded as naturally as it could be in a voxel grid or a true volumetric neural field.

\section{Verdict}

The preceding sections have surveyed a range of 3D representation paradigms, from traditional geometric primitives like meshes and voxels to modern, self-learning implicit and explicit neural models. 

In Table \ref{tab:comparison_table} we have listed estimated ratings for each representation archetype based on the evaluation criteria outlined in Section \ref{eval_criteria} and illustrated their differences in Fig. \ref{fig:comparison}. Whilst a true comparison would depend highly on implementation specifics and parameters, the overall trend reveals that the visual fidelity of modern methods comes at a cost in performance in other criteria. 

As such, we believe a praxis-oriented implementation requires moving away from a monolithic, single-purpose representation and toward a practical, hybrid system. In this section, we explore tentative ideas and concepts for how such a hybrid system may be structured.

\begin{table}[h!]
\centering
\caption{A Comparison of 3D Representation Methods}
\label{tab:comparison_table}
\resizebox{\textwidth}{!}{%
\begin{tabular}{lcccccccc}
\toprule
\textbf{Metric} & \textbf{Point Cloud} & \textbf{Voxel Grid} & \textbf{Mesh} & \textbf{Neural Field} & \textbf{3DGS} \\
\midrule
Write Performance         & ***** & ***** & ** & * & *** \\
Memory Footprint          & * & ** & **** & ***** & ***** \\
Computational Performance & * & *** & ***** & * & * \\
Surface Fidelity (Lidar)  & ***** & *** & ***** & ** & ** \\
Visual Fidelity (Camera)  & ** & * & ** & ***** & ***** \\
Volumetric Fidelity (Misc.)& * & ***** & ** & **** & * \\
\bottomrule
\end{tabular}%
}
\end{table}

\subsection{Observations and Possible Solutions}

Based on the foregoing survey, we can make several observations on the status quo of 3D representation methodology. Below, we list each of these observations in turn, alongside possible approaches to address them.

\subsubsection{Scene Management requires Hierarchy}

Regardless of the chosen representation, enabling efficient processing and rendering of large environments will remain a challenge. While it may seem "elegant" to contain an entire environment within a single neural field or Gaussian splat, it is unlikely to be practical. Instead, a hierarchical acceleration structure, such as a Bounding Volume Hierarchy (BVH), is essential for efficient culling, instancing, and managing Levels of Detail (LODs).

\begin{figure} [ht]
   \begin{center}
   \begin{tabular}{c}
   \includegraphics[height=12.2cm]{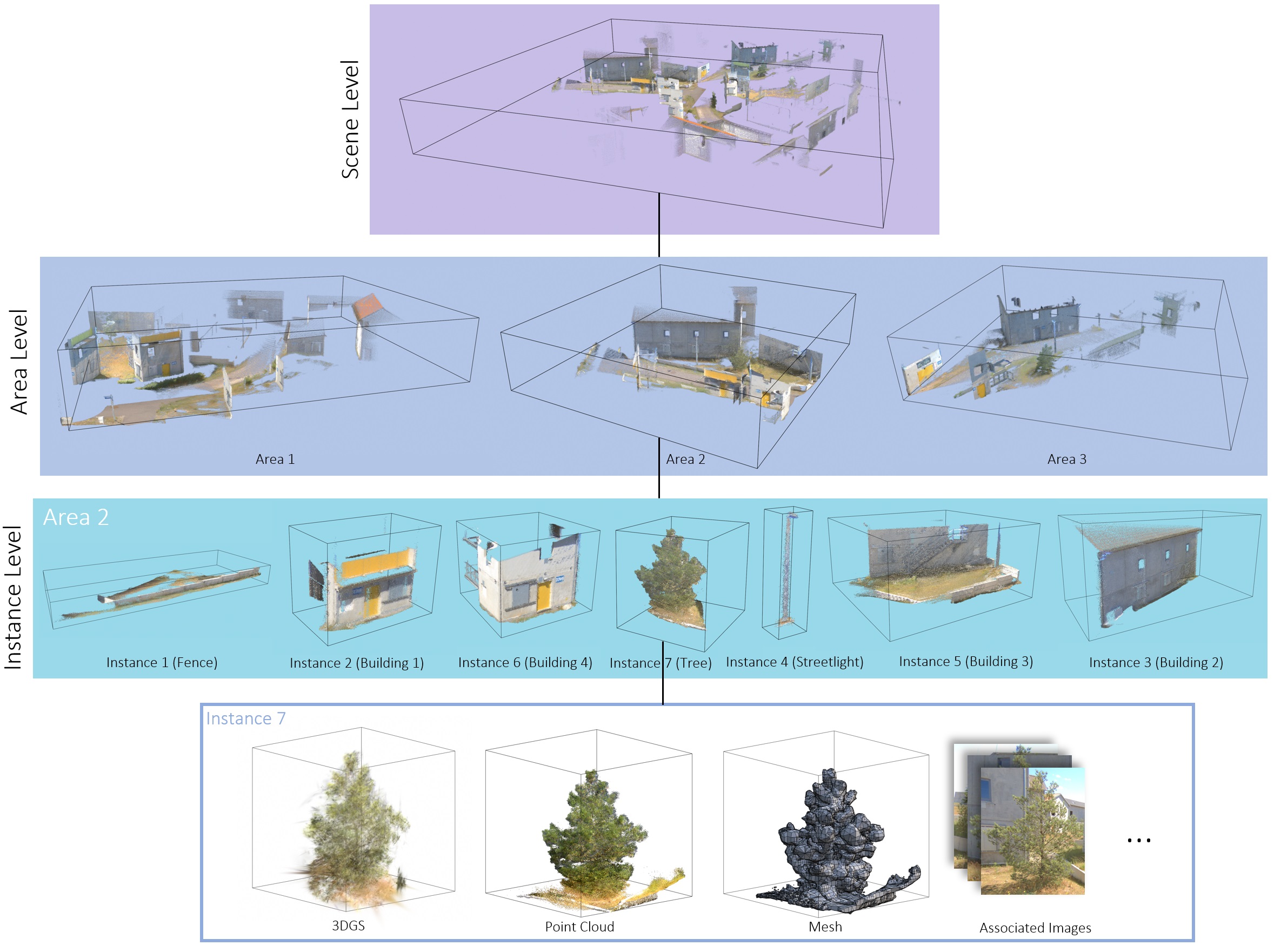}
   \end{tabular}
   \end{center}
   \caption[example] 
   { \label{fig:bvh_sample} 
   Example of a 3-level BVH applied to a scene. An area-division acts as a top-level acceleration structure, while the data is stored within instances that are trained and managed individually. The LOD of instances can be downscaled or their underlying model swapped out depending on need or distance.
   }
\end{figure}

Within such a "super-structure," the environment would consist of a hierarchy of instances with bounding volumes and transform information (see Fig. \ref{fig:bvh_sample}). The data within these instances could then be loaded and rendered dynamically on demand, or even swapped for lower-fidelity models based on distance. For instance, an object could be displayed as a high-detail Gaussian splat up close but be reduced to a simple mesh or even a basic impostor/billboard at larger distances.

This concept would also handle dynamic scenes more effectively, as instances can be moved, added, or removed with relative ease. Furthermore, annotative information such as semantic labels could be stored on a per-instance level, allowing a user to query and interact with the system through these labels, perhaps even with the aid of \textit{Large Language Models} (LLMs).

Due to their widely employed nature within game engines and graphics frameworks \cite{rtx_radiosity}, we believe some form of bounding volume hierarchy is the most viable contender for a dynamic and manageable acceleration structure.

\subsubsection{Functionality requires Geometric Grounding}

\begin{figure} [ht]
   \begin{center}
   \begin{tabular}{c}
   \includegraphics[height=9.2cm]{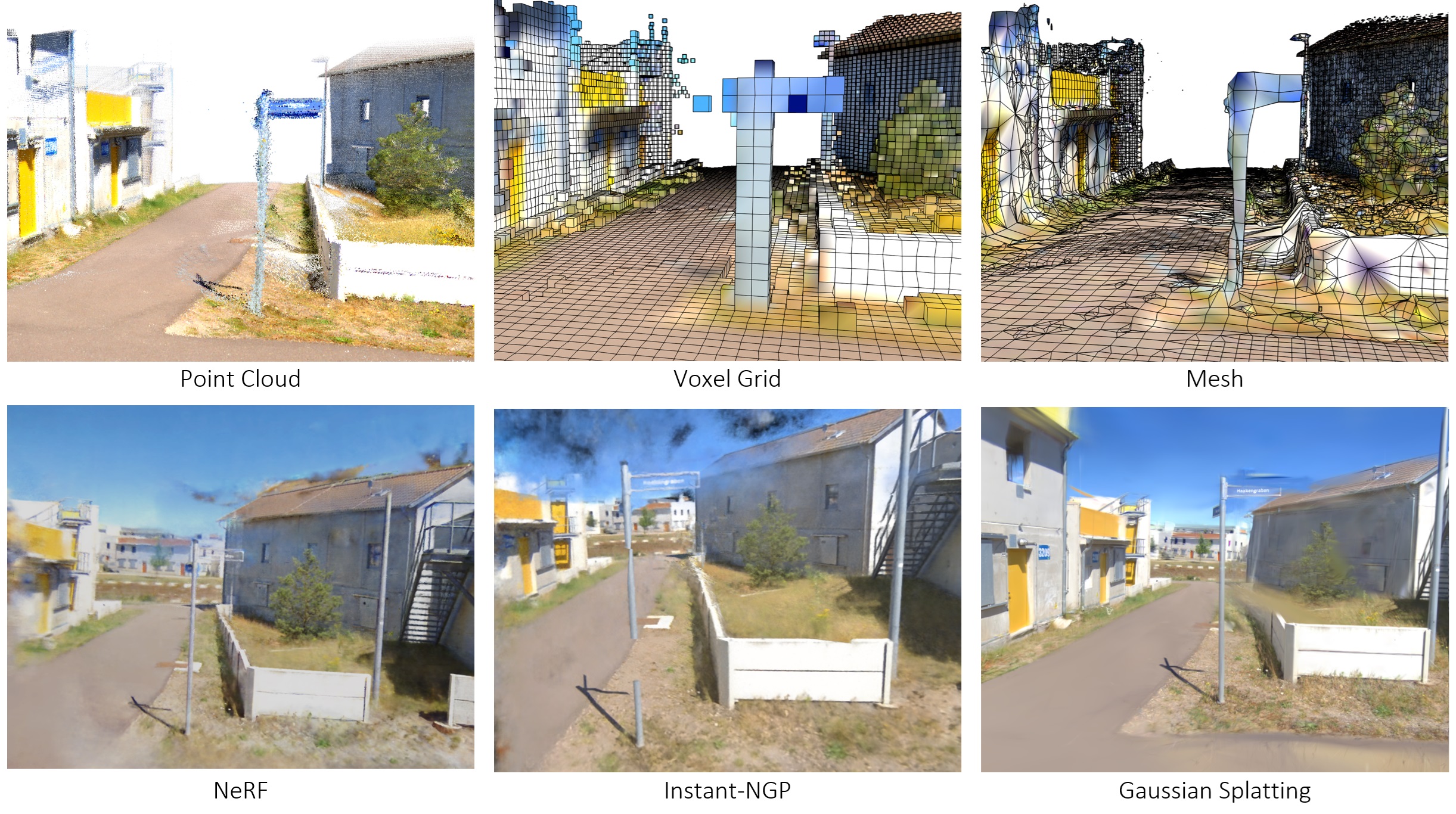}
   \end{tabular}
   \end{center}
   \caption[example] 
   { \label{fig:comparison} 
   Comparison of different representation forms of the same scene. A direct qualitative comparison is not possible, as each format uses different parameters, memory and software. But the results highlight the focus of newer methods on camera data, while more traditional models are more reliant on geometric shape.
   }
\end{figure}

Traditional models (particularly meshes) offer robust geometric guarantees and computational efficiency for tasks like physics and visibility but lack the ability to capture photorealistic appearance from images (see Fig. \ref{fig:comparison}). Conversely, modern methods like NeRFs and Gaussian Splatting excel at photorealism but are often geometrically unreliable and computationally prohibitive for tasks other than rendering \cite{sugar}.

As such, we expect that any system leveraging these technologies for more than just image synthesis would benefit greatly from an explicit geometric scaffold. This "traditional" mesh could be used for all physics, collision, and visibility calculations, while the modern representation is employed for photorealism, thus retaining the strengths of both models.

For instance, rendering a NeRF could be accelerated by first ray-tracing against an underlying mesh and then querying the NeRF only at the point of intersection. Similarly, inserting physics-based objects into a scene represented by 3DGS would be far more stable if the underlying physics calculations were performed on an encompassing mesh rather than the Gaussians themselves.

We do not expect these geometric doubles to be modeled manually, but rather to be generated by one of the various neural mesh reconstruction methods such as those outlined in Section \ref{meshes}.

\subsubsection{Fusing Sensor Modalities}

Modern approaches such as 3DGS and NeRFs primarily focus on data from camera sensors, providing great photorealistic detail but with potentially flimsy geometry. In contrast, voxel grids and point clouds make far more use of data recorded by lidar sensors, providing a more solid geometric construct but without the same level of visual detail.

A hybrid model should leverage data from both sensor types: lidar for robust shape and cameras for rich visual appearance.

\subsection{Proposal for a Hybrid Model}

Based on these observations, we propose a hybrid, instance-based representation pipeline designed for both functional utility and visual fidelity.

As a first step, the pipeline would perform instance segmentation on incoming lidar point cloud data to create the initial instances for the BVH. Each instance would be associated with relevant camera images and any generated semantic labels. This step could be performed by point-cloud segmenters like PointNet, or by methods that use complementary camera data for segmentation, such as Better Call SAL \cite{better_call_sal}, LDLS \cite{ldls}, or LIF-Seg \cite{lif-seg}.

\begin{figure} [ht]
   \begin{center}
   \begin{tabular}{c}
   \includegraphics[height=6cm]{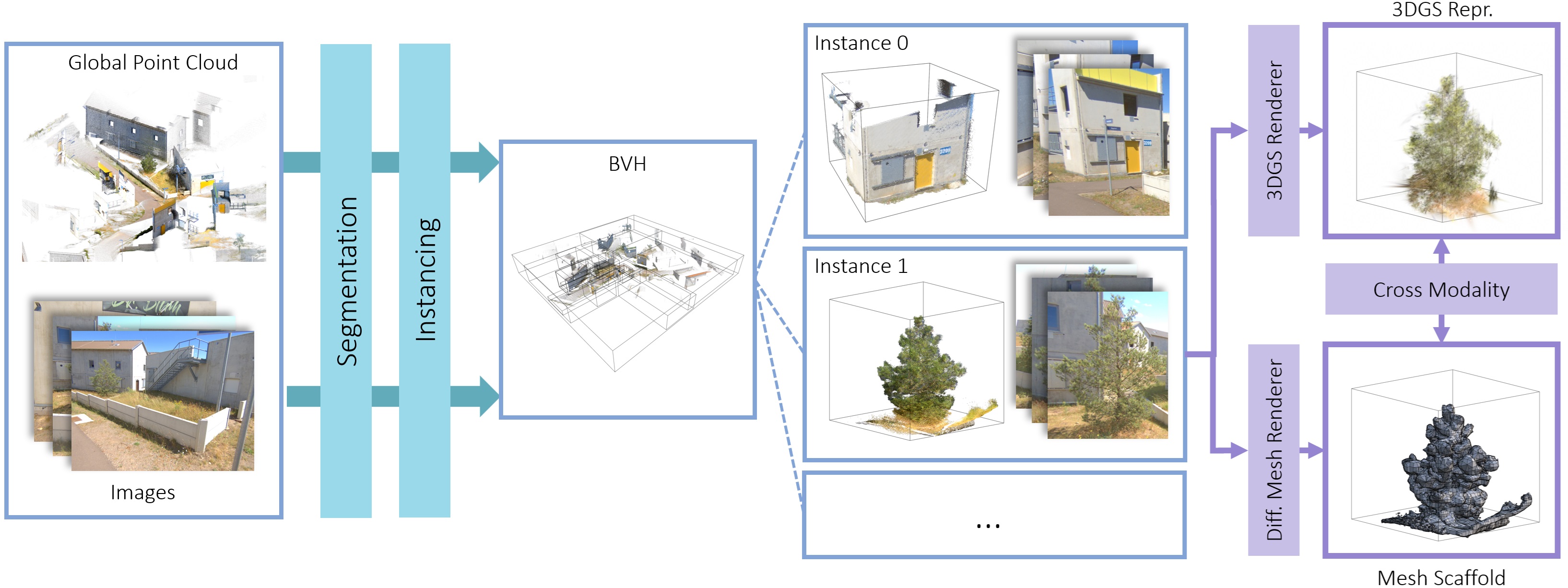}
   \end{tabular}
   \end{center}
   \caption[example] 
   { \label{fig:hybrid_arch} 
   Theoretical overview of the proposed pipeline-architecture. After data fusion, the combined point clouds and images are segmented and instanced, then a mesh scaffold and shape-constrained 3DGS representation is generated for each instance.
   }
\end{figure}

Each instance would then undergo a training process for both its realistic 3DGS representation and its geometric mesh scaffold (see Fig. \ref{fig:hybrid_arch}). This process would involve a combined loss function: a geometric loss would penalize deviations between the scaffold and the source lidar data, ensuring accuracy, while a photometric loss would optimize the properties of the Gaussians to ensure that renders match the source camera images. This dual-objective optimization ensures the final model is both geometrically sound and visually accurate.

Alternatively, the scaffold and 3DGS could be trained sequentially. One of the many surface reconstruction techniques, such as Poisson \cite{poisson} or Marching Cubes \cite{marching_cubes}, could create an initial mesh, which is then refined by a differentiable mesh renderer like Soft Rasterizer \cite{softrasterizer1} or Neural 3D Mesh Renderer \cite{neural_mesh_renderer}. This final shape would then provide positional constraints for the 3DGS representation trained thereafter. Finally, the trained Gaussians could be used to bake a high-fidelity texture for the mesh, providing a more lightweight asset.

It is important to note that this proposal provides a rough, theoretical basis. We expect there to be a number of challenges and limitations that must be addressed in a practical implementation.

\subsubsection{Related Work}

This proposed hybrid model aligns with a significant trend in recent computer graphics research that seeks to unify the benefits of classical and neural representations. Several state-of-the-art methods could serve as a foundation for this approach:

\begin{itemize}

    \item \textbf{NeuS and VolSDF:} Methods like NeuS \cite{neus} and VolSDF \cite{volsdf} are designed to learn a neural Signed Distance Function (SDF) from images using volume rendering. Because an SDF implicitly defines a surface, a high-quality mesh can easily be extracted. These models learn both a robust geometry and a neural radiance field simultaneously, providing both a functional mesh and a high-fidelity appearance model from a single training process.

    \item \textbf{SuGaR (Surface-Aligned Gaussian Splatting):} This approach is perhaps most closely aligned with our proposal. SuGaR \cite{sugar} explicitly optimizes a set of 3D Gaussians to conform to a coarse geometric mesh. It uses the mesh as a prior to regularize the Gaussians, resulting in a clean, well-defined surface while leveraging the rendering speed and quality of Gaussian Splatting. This reframes the Gaussians as an incredibly advanced neural texture for the mesh, where the mesh provides the stable geometric foundation and the Gaussians provide the fine visual detail.

\end{itemize}

\subsection{Conclusions}

Based on our evaluation of existing representation forms and the observations we were able to draw from them in the sections above, we can draw the following core conclusions:

\begin{itemize}

    \item \textit{Conclusion \#1:} While modern representations like 3DGS or NeRFs outpace traditional models in visual quality, they are often inadequate for functional tasks such as visibility calculations, route planning, or physics simulations.

    \item \textit{Conclusion \#2:} Although irregular representations like point clouds and meshes require specialized deep learning architectures, the advent of differentiable renderers makes it possible to leverage modern optimization techniques to generate and refine these traditional geometric formats.

    \item \textit{Conclusion \#3:} In our estimation, a hybrid model is the most promising path forward. Such a system would use a BVH for scene management, a "scaffold" mesh for functional operations, and a surface-aligned representation like 3DGS for visual detail, with a differentiable renderer bridging the gap between these components.

\end{itemize}

Looking ahead, the success of the hybrid model depends on continued research in key areas. Methods for simultaneous training of the mesh scaffold and surface-constrained 3DGS must be refined to optimize geometric accuracy and visual fidelity together. As components are added and challenges arise, the architecture will require ongoing adjustment. This work is not a final solution but a framework to guide the next steps toward a functional, versatile 3D battlespace visualization system.


\bibliography{report} 

@Article{social_media_analytics_mdpi,
    AUTHOR = {Sufi, Fahim},
    TITLE = {Social Media Analytics on {Russia}–{Ukraine} Cyber War with Natural Language Processing: Perspectives and Challenges},
    JOURNAL = {Information},
    VOLUME = {14},
    YEAR = {2023},
    NUMBER = {9},
    ARTICLE-NUMBER = {485},
    URL = {https://www.mdpi.com/2078-2489/14/9/485},
    ISSN = {2078-2489},
    DOI = {10.3390/info14090485}
}

@InProceedings{springer_intelligence_cycle,
    author="Biermann, Joachim",
    editor="Shahbazian, Elisa
    and Rogova, Galina
    and DeWeert, Michael J.",
    title="Understanding Military Information Processing --- An Approach to Support Intelligence in Defence and Security",
    booktitle="Harbour Protection Through Data Fusion Technologies",
    year="2009",
    publisher="Springer Netherlands",
    address="Dordrecht",
    pages="127--137",
    isbn="978-1-4020-8883-4"
}

@Article{osint_in_military,
    journal="Security and Defence Quarterly",
    issn="2300-8741",
    volume="19",
    number="2",
    year="2018",
    title="Open source intelligence ({OSINT}) as an element of military recon",
    author="Ziółkowska, Agata",
    pages="65--77",
    doi="10.5604/01.3001.0012.1474",
    url="https://doi.org/10.5604/01.3001.0012.1474"
}

@inproceedings{ai_for_c2,
      title={Artificial intelligence for decision support in command and control systems},
      author={Schubert, Johan and Brynielsson, Joel and Nilsson, Mattias and Svenmarck, Peter},
      booktitle={23rd International Command and Control Research \& Technology Symposium “Multi-Domain C},
      volume={2},
      pages={18--33},
      year={2018}
}

@article{3d_repr_survey,
      title={{3D} Representation Methods: A Survey}, 
      author={Zhengren Wang},
      year={2024},
      number={2410.06475},
      journal={arXiv},
      primaryClass={cs.CV},
      url={https://arxiv.org/abs/2410.06475}, 
}

@article{3d_gen_models_survey,
      title={Deep Generative Models on {3D} Representations: A Survey}, 
      author={Zifan Shi and Sida Peng and Yinghao Xu and Andreas Geiger and Yiyi Liao and Yujun Shen},
      year={2023},
      number={2210.15663},
      journal={arXiv},
      primaryClass={cs.CV},
      url={https://arxiv.org/abs/2210.15663}, 
}

@article{advances_3d_gen_survey,
      title={Advances in {3D} Generation: A Survey}, 
      author={Xiaoyu Li and Qi Zhang and Di Kang and Weihao Cheng and Yiming Gao and Jingbo Zhang and Zhihao Liang and Jing Liao and Yan-Pei Cao and Ying Shan},
      year={2024},
      number={2401.17807},
      journal={arXiv},
      primaryClass={cs.CV},
      url={https://arxiv.org/abs/2401.17807}, 
}

@article{survey_deeplearning_3d_repr_survey,
      title={A survey on Deep Learning Advances on Different {3D} Data Representations}, 
      author={Eman Ahmed and Alexandre Saint and Abd El Rahman Shabayek and Kseniya Cherenkova and Rig Das and Gleb Gusev and Djamila Aouada and Bjorn Ottersten},
      year={2019},
      number={1808.01462},
      journal={arXiv},
      primaryClass={cs.CV},
      url={https://arxiv.org/abs/1808.01462}, 
}

@InProceedings{DeepSDF,
author = {Park, Jeong Joon and Florence, Peter and Straub, Julian and Newcombe, Richard and Lovegrove, Steven},
title = {{DeepSDF}: Learning Continuous Signed Distance Functions for Shape Representation},
booktitle = {The IEEE Conference on Computer Vision and Pattern Recognition (CVPR)},
month = {June},
year = {2019}
}

@inproceedings{Nerf,
 title={{NeRF}: Representing Scenes as Neural Radiance Fields for View Synthesis},
 author={Ben Mildenhall and Pratul P. Srinivasan and Matthew Tancik and Jonathan T. Barron and Ravi Ramamoorthi and Ren Ng},
 year={2020},
 booktitle={Proceedings of the European Conference on Computer Vision (ECCV)},
}

@inproceedings{nerfstudio,
	title        = {Nerfstudio: A Modular Framework for Neural Radiance Field Development},
	author       = {
		Tancik, Matthew and Weber, Ethan and Ng, Evonne and Li, Ruilong and Yi, Brent
		and Kerr, Justin and Wang, Terrance and Kristoffersen, Alexander and Austin,
		Jake and Salahi, Kamyar and Ahuja, Abhik and McAllister, David and Kanazawa,
		Angjoo
	},
	year         = 2023,
	booktitle    = {ACM SIGGRAPH 2023 Conference Proceedings},
	series       = {SIGGRAPH '23}
}

@Article{gaussian_splatting,
      author       = {Kerbl, Bernhard and Kopanas, Georgios and Leimk{\"u}hler, Thomas and Drettakis, George},
      title        = {{3D} {Gaussian} Splatting for Real-Time Radiance Field Rendering},
      journal      = {ACM Transactions on Graphics},
      number       = {4},
      volume       = {42},
      month        = {July},
      year         = {2023},
      url          = {https://repo-sam.inria.fr/fungraph/3d-gaussian-splatting/}
}

@Article{mobile_mapping_sensors,
AUTHOR = {Elhashash, Mostafa and Albanwan, Hessah and Qin, Rongjun},
TITLE = {A Review of Mobile Mapping Systems: From Sensors to Applications},
JOURNAL = {Sensors},
VOLUME = {22},
YEAR = {2022},
NUMBER = {11},
ARTICLE-NUMBER = {4262},
URL = {https://www.mdpi.com/1424-8220/22/11/4262},
PubMedID = {35684883},
ISSN = {1424-8220},
DOI = {10.3390/s22114262}
}

@InProceedings{pointnet,
	author    = {Qi, Charles, R. and Su, Hao and Mo, Kaichun and Guibas, Leonidas J.},
	booktitle = {The IEEE Conference on Computer Vision and Pattern Recognition (CVPR)},
	title     = {{PointNet}: Deep Learning on Point Sets for {3D} Classification and Segmentation},
	year      = {2017},
	pages     = {77-85},
	doi       = {10.1109/CVPR.2017.16},
}

@inproceedings{pointnet_plusplus,
author = {Qi, Charles R. and Yi, Li and Su, Hao and Guibas, Leonidas J.},
title = {{PointNet++}: Deep hierarchical feature learning on point sets in a metric space},
year = {2017},
isbn = {9781510860964},
publisher = {Curran Associates Inc.},
address = {Red Hook, NY, USA},
booktitle = {Proceedings of the 31st International Conference on Neural Information Processing Systems},
pages = {5105–5114},
numpages = {10},
location = {Long Beach, California, USA},
series = {NIPS'17}
}

@inproceedings{surfels,
author = {Pfister, Hanspeter and Zwicker, Matthias and van Baar, Jeroen and Gross, Markus},
title = {Surfels: Surface elements as rendering primitives},
year = {2000},
isbn = {1581132085},
publisher = {ACM Press/Addison-Wesley Publishing Co.},
address = {USA},
url = {https://doi.org/10.1145/344779.344936},
doi = {10.1145/344779.344936},
booktitle = {Proceedings of the 27th Annual Conference on Computer Graphics and Interactive Techniques},
pages = {335–342},
numpages = {8},
series = {SIGGRAPH '00}
}

@article{point_rendering,
author = {Günther, Christian and Kanzok, Thomas and Linsen, Lars and Rosenthal, Paul},
year = {2013},
month = {06},
pages = {153},
title = {A {GPGPU}-based Pipeline for Accelerated Rendering of Point Clouds},
volume = {21},
journal = {Journal of WSCG}
}

@article{point_rendering2,
author = {Schütz, Markus and Kerbl, Bernhard and Wimmer, Michael},
title = {Rendering Point Clouds with Compute Shaders and Vertex Order Optimization},
journal = {Computer Graphics Forum},
volume = {40},
number = {4},
pages = {115-126},
doi = {https://doi.org/10.1111/cgf.14345},
url = {https://onlinelibrary.wiley.com/doi/abs/10.1111/cgf.14345},
eprint = {https://onlinelibrary.wiley.com/doi/pdf/10.1111/cgf.14345},
year = {2021}
}

@article{potree,
	title =      "Fast Out-of-Core Octree Generation for Massive Point Clouds",
	author =     "Markus Schütz and Stefan Ohrhallinger and Michael Wimmer",
	year =       "2020",
	month =      nov,
	journal =    "Computer Graphics Forum",
	volume =     "39",
	number =     "7",
	doi =        "10.1111/cgf.14134",
	publisher =  "John Wiley & Sons, Inc.",
	pages =      "1--13",
}

@ARTICLE{point_rendering_survey,
  author={Kivi, Petrus E. J. and Mäkitalo, Markku J. and Žádník, Jakub and Ikkala, Julius and Vadakital, Vinod Kumar Malamal and Jääskeläinen, Pekka O.},
  journal={IEEE Access}, 
  title={Real-Time Rendering of Point Clouds With Photorealistic Effects: A Survey}, 
  year={2022},
  volume={10},
  number={},
  pages={13151-13173},
  doi={10.1109/ACCESS.2022.3146768}
}

@article{surfel_splatting,
author = {Yifan, Wang and Serena, Felice and Wu, Shihao and \"{O}ztireli, Cengiz and Sorkine-Hornung, Olga},
title = {Differentiable surface splatting for point-based geometry processing},
year = {2019},
issue_date = {December 2019},
publisher = {Association for Computing Machinery},
address = {New York, NY, USA},
volume = {38},
number = {6},
issn = {0730-0301},
url = {https://doi.org/10.1145/3355089.3356513},
doi = {10.1145/3355089.3356513},
journal = {ACM Trans. Graph.},
month = nov,
articleno = {230},
numpages = {14}
}

@inproceedings{v3d_r2n2,
  title={{3D-R2N2}: A Unified Approach for Single and Multi-view {3D} Object Reconstruction},
  author={Choy, Christopher B and Xu, Danfei and Gwak, JunYoung and Chen, Kevin and Savarese, Silvio},
  booktitle = {Proceedings of the European Conference on Computer Vision ({ECCV})},
  year={2016}
}

@inproceedings{3dgan,
  title={Learning a probabilistic latent space of object shapes via {3D} generative-adversarial modeling},
  author={Wu, Jiajun and Zhang, Chengkai and Xue, Tianfan and Freeman, William T and Tenenbaum, Joshua B},
  booktitle={Advances in Neural Information Processing Systems},
  pages={82--90},
  year={2016}
}

@inproceedings{deepvoxels,
        author = {Sitzmann, Vincent
                  and Thies, Justus
                  and Heide, Felix
                  and Nie{\ss}ner, Matthias
                  and Wetzstein, Gordon
                  and Zollh{\"o}fer, Michael},
        title = {{DeepVoxels}: Learning Persistent {3D} Feature Embeddings},
        booktitle = {Proc. Computer Vision and Pattern Recognition (CVPR), IEEE},
        year={2019}
      }

@inproceedings{plenoxels,
  author={Fridovich-Keil, Sara and Yu, Alex and Tancik, Matthew and Chen, Qinhong and Recht, Benjamin and Kanazawa, Angjoo},
  booktitle={2022 IEEE/CVF Conference on Computer Vision and Pattern Recognition (CVPR)}, 
  title={Plenoxels: Radiance Fields without Neural Networks}, 
  year={2022},
  volume={},
  number={},
  pages={5491-5500},
  doi={10.1109/CVPR52688.2022.00542}
}

@article{instant_ngps,
    author = {Thomas M\"uller and Alex Evans and Christoph Schied and Alexander Keller},
    title = {Instant Neural Graphics Primitives with a Multiresolution Hash Encoding},
    journal = {ACM Trans. Graph.},
    issue_date = {July 2022},
    volume = {41},
    number = {4},
    month = jul,
    year = {2022},
    pages = {102:1--102:15},
    articleno = {102},
    numpages = {15},
    url = {https://doi.org/10.1145/3528223.3530127},
    doi = {10.1145/3528223.3530127},
    publisher = {ACM},
    address = {New York, NY, USA},
}

@article{voxel_cone_tracing,
author = {Crassin, Cyril and Neyret, Fabrice and Sainz, Miguel and Green, Simon and Eisemann, Elmar},
title = {Interactive Indirect Illumination Using Voxel Cone Tracing},
journal = {Computer Graphics Forum},
volume = {30},
number = {7},
pages = {1921-1930},
doi = {https://doi.org/10.1111/j.1467-8659.2011.02063.x},
url = {https://onlinelibrary.wiley.com/doi/abs/10.1111/j.1467-8659.2011.02063.x},
eprint = {https://onlinelibrary.wiley.com/doi/pdf/10.1111/j.1467-8659.2011.02063.x},
year = {2011}
}

@article{VDBPaper,
author = {Museth, Ken},
year = {2013},
month = {07},
pages = {27:1-27:22},
title = {{VDB}: High-Resolution Sparse Volumes with Dynamic Topology},
volume = {32},
journal = {ACM Trans. Graph.},
doi = {10.1145/2487228.2487235}
}

@InProceedings{neural_mesh_renderer,
  author={Kato, Hiroharu and Ushiku, Yoshitaka and Harada, Tatsuya},
  booktitle={The IEEE Conference on Computer Vision and Pattern Recognition (CVPR)}, 
  title={Neural {3D} Mesh Renderer}, 
  year={2018},
  volume={},
  number={},
  pages={3907-3916},
  doi={10.1109/CVPR.2018.00411}
}

@inproceedings{softrasterizer1,
  author={Liu, Shichen and Chen, Weikai and Li, Tianye and Li, Hao},
  booktitle={The IEEE International Conference on Computer Vision (ICCV)}, 
  title={Soft Rasterizer: A Differentiable Renderer for Image-Based {3D} Reasoning}, 
  year={2019},
  volume={},
  number={},
  pages={7707-7716},
  doi={10.1109/ICCV.2019.00780}
}

@article{softrasterizer2,
  title={A General Differentiable Mesh Renderer for Image-based {3D} Reasoning},
  author={Liu, Shichen and Li, Tianye and Chen, Weikai and Li, Hao},
  journal={IEEE Transactions on Pattern Analysis and Machine Intelligence},
  year={2020},
  publisher={IEEE}
}

@article{meshcnn,
  title={{MeshCNN}: A Network with an Edge},
  author={Hanocka, Rana and Hertz, Amir and Fish, Noa and Giryes, Raja and Fleishman, Shachar and Cohen-Or, Daniel},
  journal={ACM Transactions on Graphics (TOG)},
  volume={38},
  number={4},
  pages = {90:1--90:12},
  year={2019},
  publisher={ACM}
}

@article{rtx_radiosity,
      title={Hardware Acceleration of Progressive Refinement Radiosity using {Nvidia} {RTX}}, 
      author={Benjamin Kahl},
      year={2023},
      number={2303.14831},
      journal={arXiv},
      primaryClass={cs.GR},
      url={https://arxiv.org/abs/2303.14831}, 
}

@article{surface_reconstruction_survey,
author = {Berger, Matthew and Tagliasacchi, Andrea and Seversky, Lee and Alliez, Pierre and Guennebaud, Gaël and Levine, Joshua and Sharf, Andrei and Silva, Cláudio},
year = {2017},
month = {03},
pages = {301-329},
title = {A Survey of Surface Reconstruction from Point Clouds},
volume = {36},
journal = {Computer Graphics Forum},
doi = {10.1111/cgf.12802}
}

@inproceedings{occupancy_networks,
  author={Mescheder, Lars and Oechsle, Michael and Niemeyer, Michael and Nowozin, Sebastian and Geiger, Andreas},
  booktitle={The IEEE Conference on Computer Vision and Pattern Recognition (CVPR)}, 
  title={Occupancy Networks: Learning {3D} Reconstruction in Function Space}, 
  year={2019},
  volume={},
  number={},
  pages={4455-4465},
  doi={10.1109/CVPR.2019.00459}
}

@InProceedings{deep_sdf,
  author={Park, Jeong Joon and Florence, Peter and Straub, Julian and Newcombe, Richard and Lovegrove, Steven},
  booktitle={The IEEE Conference on Computer Vision and Pattern Recognition (CVPR)}, 
  title={{DeepSDF}: Learning Continuous Signed Distance Functions for Shape Representation}, 
  year={2019},
  volume={},
  number={},
  pages={165-174},
  doi={10.1109/CVPR.2019.00025}
}

@inproceedings{attention_nlp,
author = {Vaswani, Ashish and Shazeer, Noam and Parmar, Niki and Uszkoreit, Jakob and Jones, Llion and Gomez, Aidan N. and Kaiser, \L{}ukasz and Polosukhin, Illia},
title = {Attention is all you need},
year = {2017},
isbn = {9781510860964},
publisher = {Curran Associates Inc.},
address = {Red Hook, NY, USA},
booktitle = {Proceedings of the 31st International Conference on Neural Information Processing Systems},
pages = {6000–6010},
numpages = {11},
location = {Long Beach, California, USA},
series = {NIPS'17}
}

@article{fourier_rff,
    title={Fourier Features Let Networks Learn High Frequency Functions in Low Dimensional Domains},
    author={Matthew Tancik and Pratul P. Srinivasan and Ben Mildenhall and Sara Fridovich-Keil and Nithin Raghavan and Utkarsh Singhal and Ravi Ramamoorthi and Jonathan T. Barron and Ren Ng},
    journal={NeurIPS},
    year={2020}
}

@article{gaussian_lod,
  title={{LODGE}: Level-of-Detail Large-Scale {G}aussian Splatting with Efficient Rendering}, 
  author={Jonas Kulhanek and Marie-Julie Rakotosaona and Fabian Manhardt and Christina Tsalicoglou and Michael Niemeyer and Torsten Sattler and Songyou Peng and Federico Tombari},
  year={2025},
  number = {2505.23158},
  volume = {},
  journal={arXiv},
}

@article{gaussian_lod2,
      title={A {LoD} of {Gaussians}: Unified Training and Rendering for Ultra-Large Scale Reconstruction with External Memory}, 
      author={Felix Windisch and Lukas Radl and Thomas Köhler and Michael Steiner and Dieter Schmalstieg and Markus Steinberger},
      year={2025},
      number={2507.01110},
      journal={arXiv},
      primaryClass={cs.GR},
      url={https://arxiv.org/abs/2507.01110}, 
}

@article{lif-seg,
      title={{LIF-Seg}: {LiDAR} and Camera Image Fusion for {3D} {LiDAR} Semantic Segmentation}, 
      author={Lin Zhao and Hui Zhou and Xinge Zhu and Xiao Song and Hongsheng Li and Wenbing Tao},
      year={2021},
      number={2108.07511},
      journal={arXiv},
      primaryClass={cs.CV},
      url={https://arxiv.org/abs/2108.07511}, 
}

@inproceedings{better_call_sal,
author = {O\v{s}ep, Aljo\v{s}a and Meinhardt, Tim and Ferroni, Francesco and Peri, Neehar and Ramanan, Deva and Leal-Taix\'{e}, Laura},
title = {Better Call {SAL}: Towards Learning to Segment Anything in Lidar},
year = {2024},
isbn = {978-3-031-72932-4},
publisher = {Springer-Verlag},
address = {Berlin, Heidelberg},
url = {https://doi.org/10.1007/978-3-031-72933-1_5},
doi = {10.1007/978-3-031-72933-1_5},
booktitle = {Computer Vision – ECCV 2024: 18th European Conference, Milan, Italy, September 29–October 4, 2024, Proceedings, Part XXXIX},
pages = {71–90},
numpages = {20},
location = {Milan, Italy}
}

@article{ldls,
author = {Wang, Brian and Chao, Wei-Lun and Wang, Yan and Hariharan, Bharath and Weinberger, Kilian and Campbell, Mark},
year = {2019},
pages = {},
journal = {arXiv},
number = {1910.13955},
title = {{LDLS}: {3-D} Object Segmentation Through Label Diffusion From {2-D} Images},
doi = {10.48550/arXiv.1910.13955}
}

@inproceedings{poisson,
author = {Kazhdan, Michael and Bolitho, Matthew and Hoppe, Hugues},
title = {Poisson surface reconstruction},
year = {2006},
isbn = {3905673363},
publisher = {Eurographics Association},
address = {Goslar, DEU},
booktitle = {Proceedings of the Fourth Eurographics Symposium on Geometry Processing},
pages = {61–70},
numpages = {10},
location = {Cagliari, Sardinia, Italy},
series = {SGP '06}
}

@article{marching_cubes,
author = {Lorensen, William E. and Cline, Harvey E.},
title = {Marching cubes: A high resolution {3D} surface construction algorithm},
year = {1987},
issue_date = {July 1987},
publisher = {Association for Computing Machinery},
address = {New York, NY, USA},
volume = {21},
number = {4},
issn = {0097-8930},
url = {https://doi.org/10.1145/37402.37422},
doi = {10.1145/37402.37422},
journal = {SIGGRAPH Comput. Graph.},
month = aug,
pages = {163–169},
numpages = {7}
}

@article{neus,
      title={{NeuS}: Learning Neural Implicit Surfaces by Volume Rendering for Multi-view Reconstruction}, 
      author={Peng Wang and Lingjie Liu and Yuan Liu and Christian Theobalt and Taku Komura and Wenping Wang},
	  journal={NeurIPS},
      year={2021}
}

@inproceedings{volsdf,
	  title={Volume rendering of neural implicit surfaces},
	  author={Yariv, Lior and Gu, Jiatao and Kasten, Yoni and Lipman, Yaron},
	  booktitle={Thirty-Fifth Conference on Neural Information Processing Systems},
	  year={2021}
	}

@inproceedings{sugar,
  author={Gu{\'e}don, Antoine and Lepetit, Vincent},
  booktitle={Conference on Computer Vision and Pattern Recognition (CVPR)}, 
  title={{SuGaR}: Surface-Aligned {Gaussian} Splatting for Efficient {3D} Mesh Reconstruction and High-Quality Mesh Rendering}, 
  year={2024},
  volume={},
  number={},
  pages={5354-5363},
  doi={10.1109/CVPR52733.2024.00512}
}

@misc{Wintak,
  title = {{WinTAK} - {Team Awareness Kit}, {Tactical Assault Kit}},
  howpublished = {\url{https://www.civtak.org/tag/wintak/}}
}

@inproceedings{surfelnerf,
  author={Gao, Yiming and Cao, Yan-Pei and Shan, Ying},
  booktitle={Conference on Computer Vision and Pattern Recognition (CVPR)}, 
  title={{SurfelNeRF}: Neural Surfel Radiance Fields for Online Photorealistic Reconstruction of Indoor Scenes}, 
  year={2023},
  volume={},
  number={},
  pages={108-118},
  doi={10.1109/CVPR52729.2023.00019}
}

@article{modissa,
author = {Borgmann, Björn and Schatz, Volker and Hammer, Marcus and Hebel, Marcus and Arens, Michael and Stilla, Uwe},
year = {2021},
month = {06},
pages = {F50-F65},
title = {{MODISSA}: a multipurpose platform for the prototypical realization of vehicle-related applications using optical sensors},
volume = {60},
journal = {Applied Optics},
doi = {10.1364/AO.423599}
}

@misc{stanford,
    title     = {The {Stanford} {3D} Scanning Repository},
    note = "URL: \url{http://graphics.stanford.edu/data/3Dscanrep/}",
}

@inproceedings{colmap1,
    author={Sch\"{o}nberger, Johannes Lutz and Frahm, Jan-Michael},
    title={Structure-from-Motion Revisited},
    booktitle={Conference on Computer Vision and Pattern Recognition (CVPR)},
    year={2016},
}

@inproceedings{colmap2,
    author={Sch\"{o}nberger, Johannes Lutz and Zheng, Enliang and Pollefeys, Marc and Frahm, Jan-Michael},
    title={Pixelwise View Selection for Unstructured Multi-View Stereo},
    booktitle={European Conference on Computer Vision (ECCV)},
    year={2016},
}
\bibliographystyle{spiebib} 

\end{document}